\documentclass[
reprint,
 amsmath,
 amssymb,
 aps,
 pra,
]{revtex4-2}

\usepackage{graphicx}
\usepackage{dcolumn}
\usepackage{bm}
\usepackage[hidelinks]{hyperref}

\usepackage{xcolor}
\usepackage{overpic}
\usepackage{comment}
\usepackage{dsfont}
\usepackage{physics}

\usepackage{cleveref}
\crefname{equation}{Eq.}{Eqs.}
\Crefname{equation}{Equation}{Equations}
\crefname{figure}{Fig.}{Figs.}
\Crefname{figure}{Figure}{Figures}
\crefname{section}{Sec.}{Secs.}
\Crefname{section}{Section}{Sections}
\crefname{appendix}{Appendix}{Appendices}
\Crefname{appendix}{Appendix}{Appendices}

\renewcommand{\d}{\mathrm{d}}
\newcommand{\ee}{\mathrm{e}}
\newcommand{\ii}{\mathrm{i}}

\newcommand{\crefs}[1]{Figs.~\ref{#1}}
\newcommand{\Crefs}[1]{Figures~\ref{#1}}

\renewcommand{\Tr}[1]{\mathrm{Tr}\left[#1\right]}

\begin{document}
\title{Quantum synchronization through the interference blockade}

\author{Tobias Kehrer}
\author{Tobias Nadolny}
\author{Christoph Bruder}
\affiliation{Department of Physics, University of Basel, Klingelbergstrasse 82, CH-4056 Basel, Switzerland}

\date{\today}

\begin{abstract}
Synchronization manifests itself in oscillators adjusting their frequencies and phases with respect to an external signal or another oscillator. In the quantum case, new features appear such as destructive interferences that can result in the suppression of phase locking. A three-level (spin-1) oscillator with equal gain and damping rates and subject to an external drive does not exhibit any 1:1 phase locking but 2:1 phase locking, i.e., its phase distribution features two maxima. This bistable locking at two opposite phases is a signature of the quantum interference synchronization blockade. An analogous behavior was found for two identical coupled spin-1 oscillators.
In this work, we consider two coupled spin-1 oscillators and a drive applied to the first spin. This leads to two interference blockades between the drive and the first spin as well as between both spins. Although both interference blockades persist for strong drive and coupling strengths, remarkably, the undriven spin does show a 1:1 phase locking to the external drive. The magnitude of the locking is proportional to the drive strength if the drive strength is small. In other words, the undriven oscillator synchronizes to the external drive through both interference blockades while the blockades persist. For a chain of three coupled spin-1 oscillators, we find synchronization between the first and third spins mediated via the blockaded, second spin.
\end{abstract}

\maketitle

\noindent

\section{Introduction}
Synchronization is observed in many different domains of the natural and life sciences \cite{10.1063/5.0026335,10.1086/394562,2005Natur.438...43S,CellCycleSynch}. It occurs in systems of coupled limit-cycle oscillators and is characterized by the adjustment of oscillation frequencies to a common frequency or the emergence of maxima in the phase distributions. Synchronization has been thoroughly studied in the context of classical nonlinear dynamics \cite{RevModPhys.77.137,Synch_Pikovsky,Synch_Strogatz,10.1143/ptp/86.6.1159}.

Recently, there has been a lot of activity in the study of synchronization in quantum systems, e.g., quantum limit-cycle oscillators \cite{PhysRevE.102.042213,PhysRevResearch.3.013130} implemented as quantum harmonic oscillators \cite{PhysRevLett.111.073603,Synch_vdP_Lee,Synch_vdP_Walter,PhysRevResearch.5.023021} or few-level quantum oscillators \cite{Synch_Drive,Synch_Entanglement,PhysRevA.101.062104} subject to incoherent gain and damping. Observations of quantum synchronization have been reported in several experimental setups such as cold atoms \cite{PhysRevLett.125.013601}, nuclear spins \cite{PhysRevA.105.062206}, trapped ions \cite{PhysRevResearch.5.033209}, and superconducting qubits \cite{PhysRevResearch.2.023026}.

A three-level quantum system in which one of the three states is stabilized by incoherent gain and damping processes has been established as a minimal quantum limit-cycle oscillator. Subject to an external drive, this spin-1 oscillator aligns its phase with the drive signal. The magnitude of this so-called 1:1 phase locking is proportional to the drive strength.
If the gain and damping rates are equal, an \textit{interference blockade} emerges leading to a complete suppression of 1:1 phase locking \cite{Synch_Drive}. In this case, the oscillator tends to align its phase in one of two positions: in phase or opposite the phase of the drive. This corresponds to 2:1 phase locking. In this work, we will use the term $n$:1 phase locking if the phase distribution of an oscillator exhibits $n$ maxima corresponding to multistable locking.
A similar effect is observed for the synchronization of two identical coupled spins 1, i.e., the absence of 1:1 phase locking and the presence of 2:1 phase locking \cite{PhysRevA.99.043804}.
Interference blockades \cite{PhysRevA.108.022216} are not the only type of blockades that have been studied in systems of quantum oscillators, for another example see \cite{PhysRevLett.118.243602}. 

\begin{figure}[t]
    \begin{overpic}[width=8.6cm]
        {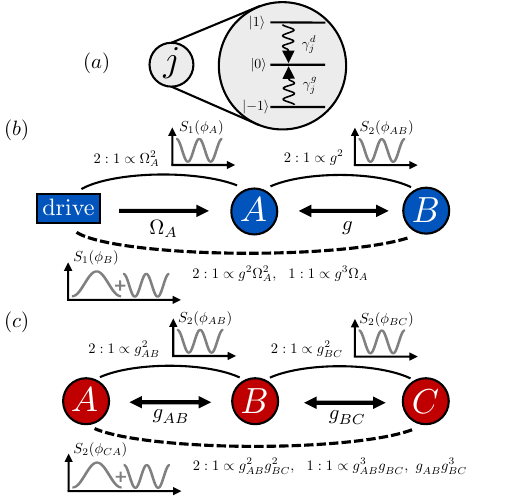}
    \end{overpic}
    \caption{Schematic of the model. (a) Each minimal quantum limit cycle oscillator labeled $A$, $B$, and $C$ consists of three spin-1 states $\ket{-1}$, $\ket{0}$, and $\ket{1}$. It is subject to two independent gain and damping processes, with rates $\gamma^g_j$ and $\gamma^d_j$, that drive the population towards state $\ket{0}$. (b) Two spins $A$ and $B$ are coupled at strength $g$ as discussed in \cref{sec:two_spin1}. Spin $A$ is furthermore driven by an external drive at rate $\Omega_A$. (c) Chain of three coupled spins without drive, see \cref{sec:three_spin1}. The insets in (b) and (c) qualitatively show the resulting phase locking of the spins. Due to blockades, 1:1 phase locking vanishes. Solid arcs denote second-order effects leading to 2:1 phase locking. Dashed arcs denote fourth-order effects leading to 1:1 and 2:1 phase locking between not directly coupled oscillators.}
    \label{fig:schematics}
\end{figure}

In this work, we first consider a drive applied to one of two coupled spin-1 oscillators. In the parameter regime of equal gain and damping rates, see \cref{fig:schematics}(b), both spins are blockaded: there is no 1:1 phase locking of the driven spin to the drive as well as no 1:1 phase locking between both spins. They both align in and out of phase corresponding to 2:1 phase locking.
Remarkably, the undriven spin \textit{does} exhibit 1:1 phase locking to the external drive. In other words, the undriven oscillator synchronizes to the external drive \textit{through} both (drive-spin and spin-spin) interference blockades \textit{without} lifting them. The locking strength is linear in the drive strength and of third order in the coupling strength. The second system that we study is a chain of three coupled spin-1 oscillators. An unexpected 1:1 phase locking, in analogy to the two-spin case, is found between the first and last spin, see \cref{fig:schematics}(c). However, the central spin mediating this locking is itself not 1:1 phase locked to any of the other two spins.

The paper is structured as follows. In \cref{sec:model}, we define the Lindblad master equation of our systems and the measure of quantum synchronization we will use. In \cref{sec:two_spin1}, we study the behavior of two spin-1 oscillators in and outside the interference blockades. In \cref{sec:three_spin1}, we analyze a system of three coupled spins 1. We summarize our results and conclude in \cref{sec:conclusion}.

\section{Quantifying Quantum Synchronization}\label{sec:model}
We consider a model of coupled spin-1 oscillators subject to gain and damping processes. The system is described by the following Lindblad master equation
\begin{align}
    \frac{\d}{\d t}\rho &= \mathcal{L}(\rho)= -\ii[H,\rho] + \sum_j \mathcal{L}_j(\rho)\,,\label{eq:EOM1}\\
    \mathcal{L}_j(\rho) &= \frac{\gamma^g_j}{2}\mathcal{D}[S^+_j S^z_j](\rho) + \frac{\gamma^d_j}{2}\mathcal{D}[S^-_j S^z_j](\rho)\,,
\end{align}
where the Hamiltonian $H$ encodes a coherent drive and spin-spin interactions and will be specified later in \cref{eq:Ham_two_spin1,eq:Ham_three_spin1}. Both incoherent processes are combined in the superoperator $\mathcal{L}_j$ and provide limit-cycle stabilization of the $j$th spin, cf.~black arrows in \cref{fig:schematics}(a). The gain and damping rates of the $j$th spin are denoted by $\gamma^g_j$ and $\gamma^d_j$ and we choose $S^z = \dyad{1}{1}-\dyad{-1}{-1}$ and $S^\pm = \sqrt{2}(\dyad{\pm 1}{0}+\dyad{0}{\mp 1})$. We use the standard notation $\mathcal{D}[L](\rho)=L\rho L^\dagger-(L^\dagger L\rho+\rho L^\dagger L)/2$. The steady state for $H=0$ is the product state $\rho^{(0)}=\dyad{0,0}{0,0}$.

A previous work shows that quantum synchronization of a single spin-1 oscillator to an external resonant drive is observed if $\gamma^g_j\neq\gamma^d_j$ \cite{Synch_Drive}. For two resonant spin-1 oscillators, quantum synchronization occurs if $\gamma^d_i+\gamma^g_j\neq\gamma^g_i+\gamma^d_j$ \cite{Synch_Entanglement}. In those works, quantum synchronization is defined as an effect that is linear in the drive strength or the interaction strength, respectively. If the rate conditions mentioned above are violated, only higher-order synchronization can be observed, i.e., the system is in the quantum interference synchronization blockade.

A variety of measures to quantify the degree of quantum synchronization has been proposed in the literature \cite{PhysRevLett.111.073603,Weiss_2016,phase_dist_Hush,Synch_Drive}. For $N$ spin-1 oscillators, we choose the synchronization measure $S_N(\vec{\phi}\,)$ defined in \cite{Synch_Entanglement},
\begin{align}
    S_N(\vec{\phi}\,) =& \left(\frac{3}{4\pi}\right)^N \int\limits_0^\pi\d\theta_1 \sin(\theta_1)\dots\nonumber\\
    &\times \int\limits_0^\pi\d\theta_N\sin(\theta_N) \bra{\vec{\theta}, \vec{\phi}}\rho\ket{\vec{\theta}, \vec{\phi}} - \frac{1}{(2\pi)^N}\,,\label{eq:Sn1}
\end{align}
where
\begin{align}
    \ket{\vec{\theta}, \vec{\phi}} &= \bigotimes_j \exp(-\ii\phi_j S^z)\exp(-\ii\theta_j S^y)\ket{1,1}\,.
\end{align}
This measure is a probability distribution of phases $\phi_j$ of each oscillator $j$ that are defined by projections of the density matrix to spin coherent states $\ket{\vec{\theta}, \vec{\phi}}$, where $\ket{S,m_S}=\ket{1,1}$ is the extremal spin-1 state. Using $S_N$, we will calculate probability distributions of relative phase angles as marginals by integrating over global phases, see, e.g., \cref{eq:SNexamples}. If the synchronization measure is flat, there is no phase preference, i.e., no synchronization in the system. Maxima of $S_N$ are related to locking of the oscillator phases. In \cref{sec:synch_measures}, we show that $S_N$ can be expressed as expectation values of powers of the spin ladder operators $S^\pm_j$, see \cref{eq:c1def,eq:SnviaOp}.
In particular, we find that the phase distributions can be written as
\begin{align}
    S_1(\phi_j) &= 2(m_j^{(1)} \cos(\phi_j) + m_j^{(2)} \cos(2\phi_j) )\,,\nonumber\\
    S_2(\phi_{ij}) &= \int\limits_0^{2\pi}\d\phi\,S_2(\phi_{ij}+\phi,\phi)\nonumber\\
    &=2(m_{ij}^{(1)} \cos(\phi_{ij}) + m_{ij}^{(2)} \cos(2\phi_{ij}) )\,,\label{eq:SNexamples}
\end{align}
where $\phi_{ij}=\phi_i-\phi_j$ is the relative phase of two oscillators $i$ and $j$. Here, we define the moments
\begin{align}
    m^{(n)}_j &= \langle (S^+_j)^n\rangle\times\begin{cases}\frac{3}{32}&n=1\,,\\\frac{1}{8\pi}&n=2\,,\\\end{cases}\label{eq:m1def}\\
    m^{(n)}_{ij} &= \langle (S^+_i S^-_j)^n\rangle\times\begin{cases}\frac{9\pi}{512}&n=1\,,\\\frac{1}{32\pi}&n=2\,,\\\end{cases}\label{eq:m2def}
\end{align}
where the label $n$ corresponds to $n$:1 phase locking and equals the number of maxima in the synchronization measure. Thus, these moments are linked to the Fourier coefficients of the phase distributions and we will use them to quantify synchronization.

\section{Two spins and a drive}\label{sec:two_spin1}
In this section, we consider two coherently coupled spins 1 labeled $A$ and $B$. A resonant coherent drive with strength $\Omega_A$ acts on spin $A$, see \cref{fig:schematics}(b). The system is described by \cref{eq:EOM1} with the Hamiltonian in the rotating frame of the drive
\begin{align}
    H =&  \frac{\Omega_A}{2}S^+_A + \frac{g}{2} S^+_A S^-_B + \text{H.c.}\,,\label{eq:Ham_two_spin1}
\end{align}
where $g$ denotes the strength of the coherent coupling. We choose both $\Omega_A$ and $g$ to be positive. Note that both spins are assumed to be in resonance with the coherent drive, i.e., the frequency of the external drive is chosen to match exactly the level spacing of the spins.

\subsection{In the interference blockade}\label{sec:intheblockade}
To study two spins 1 in the quantum interference synchronization blockade, we set the gain and damping rates $\gamma^g_A=\gamma^d_A=\gamma^g_B=\gamma^d_B=\gamma$ to be equal.
We expand the steady state $\rho_\text{ss} = \sum_{n=0}^\infty \epsilon^n \rho^{(n)}$ of \cref{eq:EOM1} in powers of $\epsilon$ for the small Hamiltonian $\epsilon H$ of \cref{eq:Ham_two_spin1}. It fulfills
\begin{align}
    \sum_j\mathcal{L}_j(\rho^{(n+1)})=\ii[H,\rho^{(n)}]\,.\label{eq:EOM1pert}
\end{align}
The synchronization measures up to fourth order in $\Omega_A$ and $g$ are
\begin{align}
    S_2(\phi_{AB}) &\approx \frac{g^2}{\pi\gamma^2} \cos(2\phi_{AB}) \left(\frac14 - 2\frac{g^2}{\gamma^2} - \frac{13}{6}\frac{\Omega_A^2}{\gamma^2}\right)\,,\label{eq:S2AB}\\
    S_1(\phi_A) &\approx \frac{\Omega_A^2}{\pi\gamma^2} \cos(2\phi_A) \left(1 - 21\frac{g^2}{\gamma^2}-8\frac{\Omega_A^2}{\gamma^2}\right)\,,\label{eq:S1A}\\
    S_1(\phi_B) &\approx \frac{5g^3\Omega_A}{2\gamma^4}\cos(\phi_B)+\frac{3g^2\Omega_A^2}{\pi\gamma^4}\cos(2\phi_B)\,,\label{eq:S1B}
\end{align}
see \cref{sec:equalrates}.
In this regime of equal gain and damping rates there is no $\cos(\phi_{A})$ and $\cos(\phi_{AB})$ contribution since both $m^{(1)}_{A}$ and $m^{(1)}_{AB}$ vanish. This is a consequence of the (drive-spin and spin-spin) interference blockades that persist for arbitrary drive and coupling strengths which means there is no 1:1 phase locking of spin $A$ to the drive and no 1:1 phase locking between spins $A$ and $B$. However, the synchronization measure $S_1(\phi_B)$ in \cref{eq:S1B} features $\cos(\phi_B)$. Hence, there is an effective first-order $\propto \Omega_A$ 1:1 phase locking of the undriven spin-1 oscillator to the drive. This 1:1 phase locking is surprising, since spin $A$ does not distinguish between the phase of the drive and its polar opposite as well as spin $B$ does not distinguish between in and out of phase locking to spin $A$. We refer to this as synchronization through the interference blockade. It is mediated via a third-order $\propto g^3$ spin-spin interaction as we will explain in more detail below. The second term in \cref{eq:S1B} denotes 2:1 phase locking of spin $B$.

In the synchronization regime where both $\Omega_A$ and $g$ are small compared to $\gamma$, the single-maximum 1:1 phase locking of the undriven spin $B$ to the drive is a small fourth-order effect. However, there is neither 1:1 phase locking of oscillator $A$ to the drive nor between oscillators $A$ and $B$ at any order in $\Omega_A$ and $g$. Both the phase distribution of oscillator $A$ and the distribution of the relative phase of oscillators $A$ and $B$ do not allow to distinguish between the phase angle of the drive and its polar opposite. For any $\Omega_A$ and $g$, only the phase distribution of oscillator $B$ uniquely reflects the phase of the drive.

This behavior can be traced back to the destructive interference of various coherences that build up. In short, even if spin $A$ does not show 1:1 phase locking to the drive, the phase of the drive is nevertheless imprinted in the coherences of the full density matrix.
Therefore, spin $B$ can exhibit 1:1 phase locking. While the contributions of the coherences to the synchronization measure of spin $A$ cancel, they do not cancel for spin $B$.

For a detailed explanation, we note that the choice of equal gain and damping rates introduces a symmetry: the master equation \cref{eq:EOM1} with the Hamiltonian \cref{eq:Ham_two_spin1} is invariant under the transformation that effectively exchanges states $\ket{j}\leftrightarrow\ket{-j}$,
\begin{align}
    S^\pm_j \to Z S^\pm_j Z^\dagger = S^\mp_j,&~~S^z_j \to Z S^z_j Z^\dagger=-S^z_j\,,\label{eq:Sym1}
\end{align}
where
\begin{align}
    Z = \exp(\ii\pi (S^x_A+S^x_B)),&~~S^x_j=(S^+_j+S^-_j)/2\,.
\end{align}
We find $\mathcal{L}(Z\rho Z^\dagger)=Z\mathcal{L}(\rho) Z^\dagger$, which leads to $\rho_\text{ss}=Z\rho_\text{ss} Z^\dagger$.
Using the invariance of the steady state under the symmetry transformation defined in \cref{eq:Sym1}, it follows that $\langle S^+_A\rangle=\langle S^-_A\rangle$ and $\langle S^+_AS^-_B\rangle=\langle S^-_AS^+_B\rangle$, hence $m^{(1)}_A\propto\langle S^+_A\rangle$ and $m^{(1)}_{AB}\propto\langle S^+_A S^-_B\rangle$ are real.
Since the master equation \cref{eq:EOM1} only consists of real parameters and $\rho^{(0)}$ is real, even orders $\rho^{(2n)}$ of the perturbation expansion of the steady state are real and odd orders $\rho^{(2n+1)}$ are purely imaginary, see \cref{eq:EOM1pert}.
At least up to fourth order in $\Omega_A$ and $g$, both $m^{(1)}_{A}$ and $m^{(1)}_{AB}$ only depend on $\rho^{(2n+1)}$, so they must be purely imaginary. Taking into account the symmetry arguments from above they must vanish in the \textit{interference blockade}: while the individual coherences do not vanish, they interfere destructively $\langle\dyad{1}{0}\otimes\mathds{1}\rangle=-\langle\dyad{0}{-1}\otimes\mathds{1}\rangle$ implying $\langle S^+_A\rangle=0$.

Spin $A$ can be intuitively interpreted as an effective drive acting on spin $B$ mediated by the spin-spin coupling.
Because of the additional coupling, $m^{(1)}_B$ depends on $\rho^{(2n)}$, and is therefore real.
The above arguments that explain the interference blockade of spin $A$ therefore do not apply, and spin $B$ is able to synchronize to the external drive.
For $m^{(1)}_B$, only the terms of order $g\Omega_A$ interfere destructively but terms of order $g^3\Omega_A$ survive which we discuss in more detail in \cref{sec:invrates}.
\begin{figure}[t]
    \includegraphics[width=8.6cm]{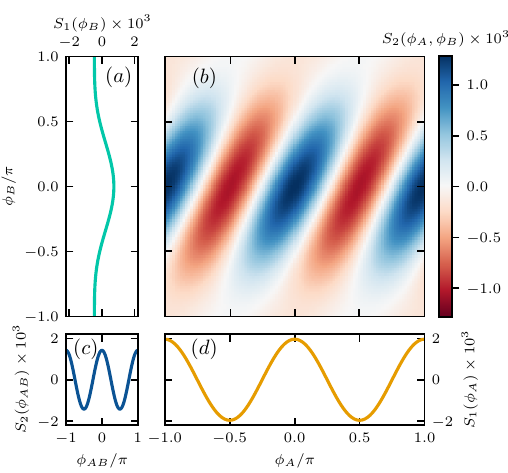}
    \caption{Synchronization measures $S_1$ and $S_2$, see \cref{eq:Sn1,eq:SNexamples}, for $\Omega_A/\gamma=0.1$ and $g/\gamma=0.15$. (b) Combined synchronization measure $S_2(\phi_A,\phi_B)$. (a),(d) Single synchronization measures $S_1(\phi_A)$ and $S_1(\phi_B)$ as marginals of (b). (c) Combined synchronization measure $S_2(\phi_{AB})$. Both $S_2(\phi_{AB})$ and $S_1(\phi_A)$ exhibit two maxima, whereas $S_1(\phi_B)$ of the undriven spin in panel (a) is characterized by only one maximum.}
    \label{fig:S2_two_spin1}
\end{figure}
\begin{figure}[t]
    \includegraphics[width=8.6cm]{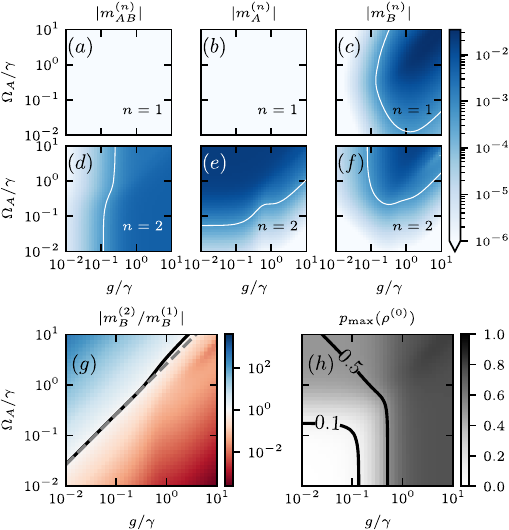}
    \caption{(a)--(f) First ($n=1$) and second ($n=2$) moments indicating one and two maxima in the corresponding synchronization measures. The white curves are contour lines of the moments at $5\times 10^{-4}$. (g) Ratio of the second and first moment of spin $B$. The black curve denotes $|m^{(2)}_B/m^{(1)}_B|=1$ and the gray dashed lines denote the corresponding theoretical prediction $\Omega_A=5\pi g/6$ based on \cref{eq:S1B}. (h) Maximum change of state populations, see \cref{eq:pmax}.}
    \label{fig:S_moments_two_spins}
\end{figure}

In \cref{fig:S2_two_spin1}, we plot the individual synchronization measures $S_1(\phi_A)$ and $S_1(\phi_B)$ as well as the combined measures $S_2(\phi_A,\phi_B)$ and $S_2(\phi_{AB}=\phi_A-\phi_B)$, that are defined in \cref{sec:synch_measures}, evaluated for the numerically exact steady state of \cref{eq:EOM1}. As expected from \cref{eq:S2AB,eq:S1A,eq:S1B}, both $S_1(\phi_A)$ and $S_2(\phi_{AB})$ show two maxima, see \crefs{fig:S2_two_spin1}(c) and (d). These two distributions imply that spin $A$ locks with two preferred phases to the drive and spin $B$ locks with two preferred phases to spin $A$. Therefore, one could naively conclude that spin $B$ also exhibits two maxima in its phase distribution. 
However, this is not true in general. \Cref{fig:S2_two_spin1}(b) shows that the maxima of the combined quantum synchronization measure lie at $(\phi_A,\phi_B)\in\lbrace (0,0), (\pi,0)\rbrace$, leading to the single maximum of $S_1(\phi_B)$, see \cref{fig:S2_two_spin1}(a).

In \cref{fig:S_moments_two_spins}, we show moments that reflect the synchronization behavior, see \cref{eq:m1def,eq:m2def}, for various drive and coupling strengths. As predicted by \cref{eq:S2AB,eq:S1A,eq:S1B}, $S_1(\phi_B)$ exhibits a first moment, see \cref{fig:S_moments_two_spins}(c). In contrast, the first moment vanishes for $S_1(\phi_A)$ and $S_2(\phi_{AB})$, see \crefs{fig:S_moments_two_spins}(a) and (b). All synchronization measures show a two-maxima contribution, see \crefs{fig:S_moments_two_spins}(d) to (f). In \cref{fig:S_moments_two_spins}(g), we plot the ratio of the second and first moment of the undriven spin $B$ indicating that $S_1(\phi_B)$ exhibits predominantly two maxima if $\Omega_A \gg g$ and one maximum if $\Omega_A \ll g$. In \cref{fig:S_moments_two_spins}(h), we show the maximum change in populations between the numerically obtained density matrix $\rho^\text{ss}$ and a reference state $\rho=\rho^{(0)}=\dyad{0,0}{0,0}$ \cite{PhysRevA.99.043804}
\begin{align}
    p_\text{max}(\rho) &= \max_{n}\big|\rho^\text{ss}_{n,n}-\rho^{}_{n,n}\big|\,.\label{eq:pmax}
\end{align}
It can be used to identify the regime of synchronization in which the limit-cycle state is only weakly perturbed, i.e., $p_\text{max}\lesssim 0.1$ which we find to be $g,\Omega_A\lesssim0.1\gamma$. In this region, the fourth-order approximation agrees with the numerical results presented in \crefs{fig:S_moments_two_spins}(a) to (g). Moreover, entanglement measures are small below $g/\gamma\lesssim0.1$, see \cref{sec:entanglement}. The relation between quantum synchronization and entanglement has been studied for, e.g., spins \cite{Synch_Entanglement,Chepelianskii2024} and harmonic oscillators \cite{PhysRevLett.111.103605,PhysRevE.89.022913,GARG2023128557}. 

Note that if the gain and damping rates are chosen such that only one of either a drive-spin or a spin-spin interference blockade exists, it does not persist up to large drive and coupling strengths. The drive-spin blockade is lifted by the spin-spin interaction and vice versa. Since in these cases $m^{(1)}_A$ and $m^{(1)}_{AB}$ are not zero, it is not surprising that also $m^{(1)}_B$ is not zero. Only when imposing both blockades simultaneously by equal gain and damping rates for all spins, as described in this section, the blockades persist.

\begin{figure}[t]
    \includegraphics[width=8.6cm]{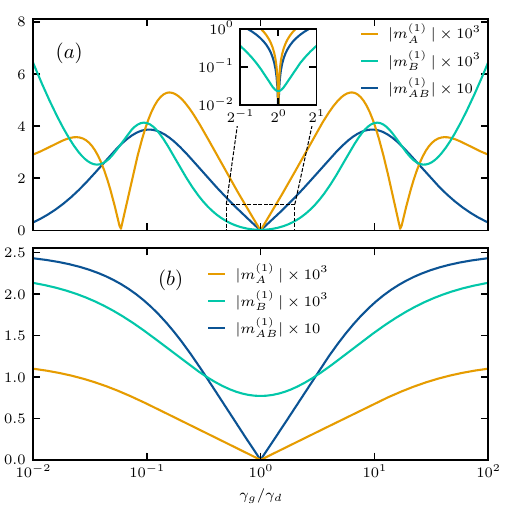}
    \caption{First moments of the individual and combined synchronization measures for $\Omega_A/(\gamma_g+\gamma_d)=10^{-3}$. (a) $g/(\gamma_g+\gamma_d)=0.05$. The inset highlights the region close to the interference blockade $\gamma_g=\gamma_d$. (b) $g/(\gamma_g+\gamma_d)=0.5$.}
    \label{fig:gblockade}
\end{figure}

\subsection{Outside the interference blockade}\label{sec:outsideblockade}
In the previous section, synchronization is blockaded perfectly. We now discuss the behavior of the two-spin system for inverted gain and damping rates $\gamma^g_A=\gamma^d_B=\gamma_g$ and $\gamma^d_A=\gamma^g_B=\gamma_d$ close to the blockade. To this end, we show $m^{(1)}_A$, $m^{(1)}_B$, and $m^{(1)}_{AB}$ in \cref{fig:gblockade}. Whenever $\gamma_g\neq\gamma_d$, the symmetry defined by \cref{eq:Sym1} is broken and the interference blockades disappear such that 1:1 drive-spin and spin-spin phase locking exist.
Nevertheless, there is a regime in which $|m^{(1)}_A| < |m^{(1)}_B|$. Its width can be estimated by expanding the ratio of the first moments of spin $A$ and spin $B$ to first order in $\gamma_g/\gamma_d-1$. This expansion can be used to approximatively solve $|m^{(1)}_A/m^{(1)}_B|=1$ by
\begin{align}
    \frac{\gamma_g}{\gamma_d} &= 1\pm \frac{20 g^3}{3 \gamma_d^3} + \mathcal{O}(g^5/\gamma_d^5) \approx 1\pm \frac{160 g^3}{3 (\gamma_g + \gamma_d)^3}\,.
\end{align}
The region in which the undriven spin $B$ exhibits a stronger 1:1 phase locking to the drive than the driven spin $A$ has an approximate width $\propto g^3/\gamma_d^3$ in terms of the ratio of gain and damping rates $\gamma_g/\gamma_d$.

In addition to the interference blockade, i.e., vanishing $m^{(1)}_A$ and $m^{(1)}_{AB}$, between spin $A$ and its drive as well as between both spins we find another synchronization blockade that is induced by the coupling. This new and additional blockade appears at roots of $m^{(1)}_A$ and $m^{(1)}_{AB}$ for values of $\gamma_g/\gamma_d$ depending on $g$, see \cref{eq:gblockadem1AB,eq:gblockadem1A1,eq:gblockadem1A2} and \cref{fig:gblockade}(a). In the interference blockade $\gamma_g=\gamma_d$, up to first order in $\Omega_A$, contributions to $m^{(1)}_{AB}$ originating from both $\dyad{0,1}{1,0}$ and $\dyad{-1,0}{0,-1}$ vanish individually, whereas terms proportional to $\dyad{0,0}{1,-1}$ and $\dyad{-1,1}{0,0}$ cancel. In the coupling-induced blockade, these coherences cancel collectively. 

The coupling-induced blockades occur for rather large coupling strengths for which the steady state of the system deviates significantly from $\rho^{(0)}$. In the regime $g\gtrsim\gamma_g+\gamma_d$ one obtains $p_\text{max}(\rho^{(\infty)})\lesssim 0.1$, i.e., the steady state is close to $\rho^{(\infty)}$.

\begin{figure}[t]
    \includegraphics[width=8.6cm]{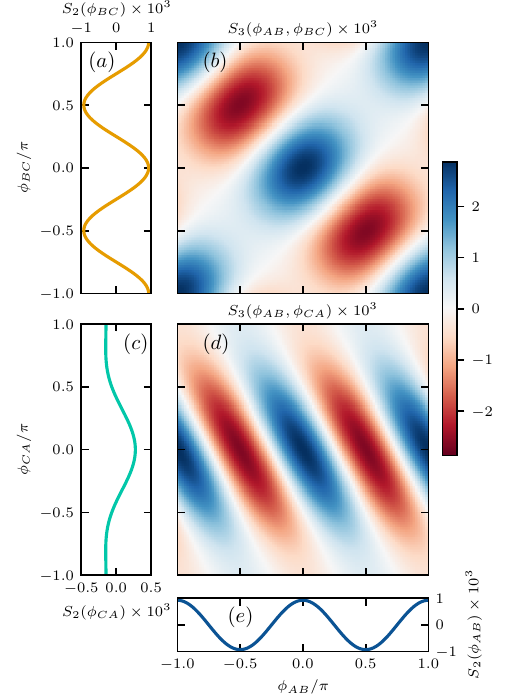}
    \caption{Synchronization measures $S_2$ and $S_3$, see \cref{eq:SNexamples,eq:S3ABBC,eq:S3ABCA}, for $g_{AB}/\gamma=g_{BC}/\gamma=0.12$. (b),(d) Combined measures for three coupled spin-1 oscillators. (a),(c),(e) Combined measures for pairs of two spins 1.}
    \label{fig:S3_three_spin1}
\end{figure}
\begin{figure}[t]
    \includegraphics[width=8.6cm]{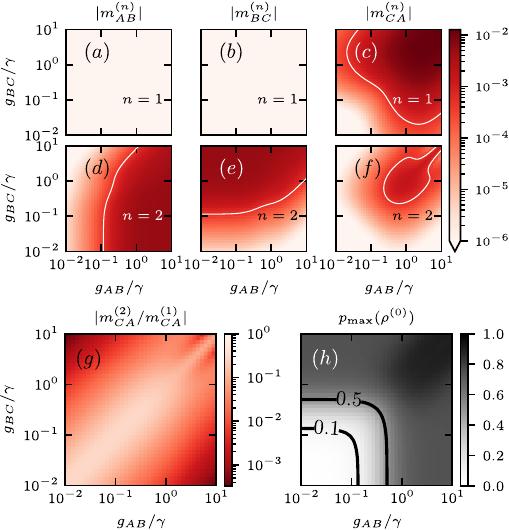}
    \caption{(a)--(f) First ($n=1$) and second ($n=2$) moments indicating one and two maxima in the corresponding synchronization measures. The white curves are contour lines of the moments at $5\times 10^{-4}$. (g) Ratio $|m^{(2)}_{CA}/m^{(1)}_{CA}|<1$ of the second and first moment of the combined measure of spins $A$ and $C$. (h) Maximum change of a state populations, see \cref{eq:pmax}.}
    \label{fig:S_moments_three_spins}
\end{figure}

\section{Three undriven spins}\label{sec:three_spin1}
We now consider a system of three undriven coupled spin-1 oscillators labeled $A$, $B$, and $C$,
\begin{align}
    H =&  \frac{g_{AB}}{2} S^+_A S^-_B + \frac{g_{BC}}{2} S^+_B S^-_C + \text{H.c.}\,,\label{eq:Ham_three_spin1}
\end{align}
where $g_{AB}$ ($g_{BC}$) is the coupling strength between spins $A$ and $B$ ($B$ and $C$). Similar to \cref{sec:intheblockade}, all gain and damping rates are set equal to $\gamma$. In \cref{fig:S3_three_spin1}, we show both synchronization measures $S_3(\phi_{AB}, \phi_{BC})$ and $S_3(\phi_{AB}, \phi_{CA})$, evaluated for the numerically exact steady state of \cref{eq:EOM1}, for three spins which are probability distributions of the relative phases between the three oscillators, see \cref{eq:S3ABBC,eq:S3ABCA}. Moreover, we present the synchronization measures $S_2$ between all three pairs of spins as marginals. As expected, $S_2(\phi_{AB})$ and $S_2(\phi_{BC})$ of both pairs of coupled spins exhibit two maxima due to the quantum interference synchronization blockade, see \crefs{fig:S3_three_spin1}(a) and (e). However, similar to the single-maximum locking of the undriven spin $B$ in \cref{fig:S2_two_spin1}(a), the synchronization measure between the spins $A$ and $C$ that are not directly coupled exhibits a single maximum in the phase difference $\phi_{CA}$, see \cref{fig:S3_three_spin1}(c). This contradicts the naive expectation that if $S_2(\phi_{AB})$ and $S_2(\phi_{BC})$ exhibit two maxima, $S_2(\phi_{CA})$ will also exhibit two maxima. In fact, the synchronization measures $S_3(\phi_{AB}, \phi_{BC})$ and $S_3(\phi_{AB}, \phi_{CA})$ exhibit maxima at $(\phi_{AB}, \phi_{BC}, \phi_{CA})\in\lbrace (0,0,0),(\pi,\pi,0) \rbrace$ revealing the true locking behavior: the phases of neighboring spins are either aligned or anti-aligned.

In analogy to \cref{fig:S_moments_two_spins}, we display relevant moments of the three-spin system in \cref{fig:S_moments_three_spins}. \Crefs{fig:S_moments_three_spins}(a) and (b) show vanishing 1:1 phase locking between directly coupled spins. In contrast, \cref{fig:S_moments_three_spins}(c) shows 1:1 phase locking between the spins $A$ and $C$ that are not directly coupled. Similar to what was found for the undriven spin $B$ discussed in \cref{sec:two_spin1}, the quantum synchronization measure between the uncoupled spins $A$ and $C$ exhibits both non-vanishing first and second moments. All synchronization measures exhibit a two-maxima contribution, see \crefs{fig:S_moments_three_spins}(d) to (f). Interestingly, in contrast to the setup of two spin-1 oscillators, \cref{fig:S_moments_three_spins}(g) shows that the first moment always dominates, i.e., $|m^{(2)}_{CA}|<|m^{(1)}_{CA}|$. In the region $g_{AB},g_{BC}\lesssim0.1\gamma$, entanglement measures are small, see \cref{fig:ent_three_spin1} in \cref{sec:entanglement}. This matches the region of $p_\text{max}\lesssim0.1$ in \cref{fig:S_moments_three_spins}(h).

\section{Conclusion}\label{sec:conclusion}
We have analyzed setups of two and three coupled spin-1 oscillators in the parameter regime of equal gain and damping rates leading to (spin-spin) quantum interference blockades between all coupled oscillators. In the case of two spins, a drive acting on spin $A$ leads to a second type of (drive-spin) quantum interference blockade. Both blockades persist for arbitrarily large drive and coupling strengths.

In the two-spin setup, the blockades manifest themselves in vanishing first moments of the quantum synchronization measure of spin $A$ as well as of the combined synchronization measure of both spins. Spin $A$ synchronizes with equal probability in and out of phase with the drive with a magnitude proportional to the square of the drive strength $\Omega_A$. Similarly, spin $B$ locks in and out of phase to spin $A$ with a magnitude proportional to the square of the coupling strength $g$. The naive expectation that spin $B$ will therefore also lock with two preferred phases to the drive fails in general. The undriven spin $B$ exhibits a 1:1 phase locking to the drive through both blockades without lifting them. The magnitude of this 1:1 phase locking is proportional to $g^3 \Omega_A$ corresponding to a first-order locking to the drive mediated via a third-order spin-spin interaction. Remarkably, the driven spin $A$ exhibits no 1:1 phase locking. If the parameters are chosen such that only one of either a drive-spin or a spin-spin interference blockade exists, it does not persist up to large drive and coupling strengths. The drive-spin blockade is lifted by the spin-spin interaction and vice versa. Only when imposing both blockades simultaneously by equal gain and damping rates for all spins, the blockades persist.

In a three-spin chain, the combined quantum synchronization measures of both pairs of directly coupled spins exhibit two maxima. However, similar to the two-spin case discussed in the previous paragraph, we observe a 1:1 phase locking behavior between the two not directly coupled spins $A$ and $C$. Analogously, this locking exists without lifting the quantum interference blockades in the other two subsystems $AB$ and $BC$.

Quantum synchronization thus provides a rich set of interesting features. Even for systems whose building blocks are the simplest possible quantum limit-cycle oscillators, unexpected properties arise like the locking of two not directly coupled spins mediated by an intermediate spin that is itself not locked. An intriguing question for the future is the study of the competition of single-maximum (indirect coupling) and two-maxima locking (direct coupling) in geometrically frustrated configurations of spin-1 oscillators.

\acknowledgments
We acknowledge financial support from the Swiss National Science Foundation individual grant (Grant No. 200020 200481). We furthermore acknowledge the use of \textsc{QuTip} \cite{QuTiP}.

\appendix

\section{Quantum Synchronization Measures}\label{sec:synch_measures}
We use the measure of quantum synchronization for single spin-$S$ oscillators introduced in \cite{Synch_Drive},
\begin{align}
    S_1(\phi) &= \int\limits_0^\pi\d\theta\,\sin(\theta)Q(\theta,\phi,\rho)-\frac{1}{2\pi}\,,\label{eq:S1def}
\end{align}
where
\begin{align}
    Q(\theta,\phi,\rho) &= \frac{2S+1}{4\pi}\bra{\theta,\phi}\rho\ket{\theta,\phi}\label{eq:Q1def}
\end{align}
is the Husimi $Q$ function of $\rho$ with respect to spin coherent states
\begin{align}
    \ket{\theta,\phi} &= \exp(-\ii\phi S^z)\exp(-\ii\theta S^y)\ket{S,S}\,,
\end{align}
and $\ket{S,m_S=S}$ is the extremal spin-$S$ state. Using the Wigner $D$ matrix \cite{Wigner1959}, we can express \cref{eq:Q1def} as
\begin{align}
    Q(\theta,\phi,\rho) &= \frac{2S+1}{4\pi}\sum_{n,m} \ee^{\ii(n-m)\phi}d^S_{n,S}(\theta)d^S_{m,S}(\theta)\rho_{n,m}\,,
\end{align}
where
\begin{align}
    d^S_{n,S}(\theta) &= \sqrt{\frac{(2S)!}{(S+n)!(S-n)!}}\cos\left(\frac{\theta}{2}\right)^{S+n}\sin\left(\frac{\theta}{2}\right)^{S-n}\,.
\end{align}
The integration over $\theta$ in \cref{eq:S1def} can be mapped to Eq.~(3.621.5) of \cite{GradshteynRyzhik}, leading to \begin{align}
    S_1(\phi) &= \Tr{c^S(\phi)\rho}-\frac{1}{2\pi} = \langle c^S(\phi)\rangle-\frac{1}{2\pi}\,,\label{eq:S1viaOp}
\end{align}
where \cite{Tan2022halfintegervs}
\begin{align}
    c^S_{n,m}(\phi) =& \ee^{\ii(n-m)\phi}\frac{2S+1}{4\pi}\int\limits_0^\pi\d\theta\,\sin(\theta)d^S_{n,S}(\theta)d^S_{m,S}(\theta) \nonumber\\
    =&\frac{\ee^{\ii(n-m)\phi}}{2\pi}\frac{\Gamma\left(1+S+\frac{n+m}{2}\right)\Gamma\left(1+S-\frac{n+m}{2}\right)}{\sqrt{(S+n)!(S-n)!(S+m)!(S-m)!}}
\end{align}
are the components of the operator $c^S(\phi)$. For spin-$1/2$ and spin-1 oscillators, we find explicit expressions for $c^S(\phi)$,
\begin{align}
    c^\frac12(\phi) &= \frac{\mathds{1}}{2\pi} + \frac18\left(\ee^{\ii\phi}S^+ + \ee^{-\ii\phi}S^-\right)\,,\label{eq:c12def}\\
    c^1(\phi) &= \frac{\mathds{1}}{2\pi} + \left(\frac{3\ee^{\ii\phi}}{32}S^+ + \frac{\ee^{\ii2\phi}}{8\pi}(S^+)^2 + \mathrm{H.c.}\right)\,.\label{eq:c1def}
\end{align}
For larger spins, the expression of $c^S$ becomes more complex, e.g., $c^\frac32(\phi)$ features terms of the form $(S^+)^2S^-$ and $S^-(S^+)^2$.

The definition of the synchronization measure \cref{eq:S1def} can be generalized to systems consisting of $N$ spin-$S$ oscillators by considering tensor products of spin coherent states, see \cite{Synch_Entanglement},
\begin{align}
    S_N(\vec{\phi}\,) =& \int\limits_0^\pi\d\theta_1 \sin(\theta_1)\dots\nonumber\\
    &\times\int\limits_0^\pi\d\theta_N \sin(\theta_N) Q(\vec{\theta}, \vec{\phi},\rho) - \frac{1}{(2\pi)^N}\,,\label{eq:Sndef}
\end{align}
where
\begin{align}
    Q(\vec{\theta}, \vec{\phi},\rho) &= \left(\frac{2S+1}{4\pi}\right)^N\bra{\vec{\theta}, \vec{\phi}}\rho\ket{\vec{\theta}, \vec{\phi}}\,,\\
    \ket{\vec{\theta}, \vec{\phi}} &= \bigotimes_{j=1}^N \exp(-\ii\phi_j S^z)\exp(-\ii\theta_j S^y)\ket{S,S}\,.
\end{align}
Due to this tensor-product structure, we can express \cref{eq:Sndef} as
\begin{align}
    S_N(\vec{\phi}\,) &= \left\langle \bigotimes_{j=1}^N c^S(\phi_j) \right\rangle - \frac{1}{(2\pi)^N}\,.\label{eq:SnviaOp}
\end{align}
In this work, we are interested in up to three spin-1 oscillators. Combining \cref{eq:c1def,eq:SnviaOp}, for a single spin 1, we obtain
\begin{align}
    S_1(\phi) &= \left\langle\frac{3}{32}\ee^{\ii\phi}S^+ + \frac{\ee^{\ii 2\phi}}{8\pi}(S^+)^2+\mathrm{H.c.}\right\rangle\,,\label{eq:S1viaOp1}
\end{align}
and for a system consisting of two spins 1,
\begin{align}
    S_2&(\phi_{AB}) = \int\limits_0^{2\pi}\d\phi_B\,S_2(\phi_{AB}+\phi_B,\phi_B)\nonumber\\
    &= \left\langle\frac{9\pi}{512}\ee^{\ii\phi_{AB}}S^+_A S^-_B +\frac{\ee^{\ii 2\phi_{AB}}}{32\pi}(S^+_A S^-_B)^2+\mathrm{H.c.}\right\rangle\,.\label{eq:S2viaOp}
\end{align}
The structure of \cref{eq:c1def} allows us to express the Fourier transform of $S_N$ as expectation values of powers of the spin-1 ladder operators $S^+_j$. Note that the coefficients of \cref{eq:S2viaOp} $9\pi/512=2\pi(3/32)^2$ and $1/32\pi=2\pi/(8\pi)^2$ are related to squares of the coefficients of \cref{eq:S1viaOp1}, where the additional factor of $2\pi$ arises from the integration over $\phi_B$. Similarly, for three spins, we obtain
\begin{align}
    S_3(\phi_{AB}, \phi_{BC}) &= \int\limits_0^{2\pi}\d\phi_B\,S_3(\phi_{AB}+\phi_B, \phi_B, \phi_B - \phi_{BC})\,,\label{eq:S3ABBC}\\
    S_3(\phi_{AB}, \phi_{CA}) &= \int\limits_0^{2\pi}\d\phi_A\,S_3(\phi_A, \phi_A-\phi_{AB}, \phi_{CA}+\phi_A)\,.\label{eq:S3ABCA}
\end{align}

\section{Strong Drive and/or Coupling}\label{sec:strong_drive_coupling}
For vanishing drive $\Omega_A=0$ or vanishing coupling $g=0$, we can solve the system of two spin-1 oscillators for arbitrary gain and damping rates analytically. 

\subsection{Inverted Gain and Damping Rates}\label{sec:invrates}
In the case of inverted gain and damping rates $\gamma^g_A=\gamma^d_B=\gamma_g$ and $\gamma^d_A=\gamma^g_B=\gamma_d$ both $m^{(1)}_A$ and $m^{(1)}_{AB}$ do not vanish. Solving the system for $\Omega_A=0$ and adding a drive that acts on spin $A$ as a small perturbation leads to the following leading-order contributions in $\Omega_A/\gamma$,
\begin{widetext}
    \begin{align}
        m^{(1)}_A &= \ii\frac{3\Omega_A}{16}\frac{\gamma_g-\gamma_d}{\gamma_g\gamma_d} \left(1 - 4 g^2\frac{(\gamma_g^2 + 4\gamma_g\gamma_d + \gamma_d^2)}{\gamma_g^2\gamma_d^2} + \mathcal{O}\left(\frac{g^4}{\gamma_d^4}\right)\right)\,,\label{eq:m1Aasym}\\
        m^{(1)}_B &= \frac{3\Omega_A g}{8\gamma_g\gamma_d} \left(\frac{(\gamma_d-\gamma_g)^2}{\gamma_g\gamma_d} + g^2 \frac{320 \gamma_g^3 \gamma_d^3 + 23 (\gamma_g^4 \gamma_d^2 + \gamma_g^2 \gamma_d^4) - 32 (\gamma_g^6 + \gamma_d^6) - 106 (\gamma_g^5 \gamma_d+\gamma_g \gamma_d^5)}{
 3 \gamma_g^3 \gamma_d^3 (2 \gamma_g + \gamma_d) (\gamma_g + 
    2 \gamma_d)}+ \mathcal{O}\left(\frac{g^4}{\gamma_d^4}\right)\right)\label{eq:m1Basym}\\
        m^{(1)}_{AB} &=
        \ii\frac{9\pi g}{256}\frac{\overbrace{2(\gamma_d-\gamma_g)g^2\gamma_g\gamma_d}^{\text{from}~\dyad{0,1}{1,0}} + \overbrace{2(\gamma_d-\gamma_g)g^2\gamma_g\gamma_d}^{\text{from}~\dyad{-1,0}{0,-1}} + (4g^2+\gamma_g\gamma_d)(\overbrace{(g^2+\gamma_g^2)\gamma_d}^{\text{from}~\dyad{0,0}{1,-1}} - \overbrace{(g^2+\gamma_d^2)\gamma_g}^{\text{from}~\dyad{-1,1}{0,0}})}{32 g^6 + \gamma_g^3\gamma_d^3 + 4 g^4 (2 \gamma_g^2 + 7 \gamma_g \gamma_d + 2 \gamma_d^2) + g^2 \gamma_g \gamma_d  (4 \gamma_g^2 + 5 \gamma_g \gamma_d + 4 \gamma_d^2)}\nonumber\\
        &=\ii\frac{9\pi g}{256}\frac{(\gamma_d-\gamma_g)(4g^4 + g^2 \gamma_g\gamma_d - \gamma_g^2\gamma_d^2)}{32 g^6 + \gamma_g^3\gamma_d^3 + 4 g^4 (2 \gamma_g^2 + 7 \gamma_g \gamma_d + 2 \gamma_d^2) + g^2 \gamma_g \gamma_d  (4 \gamma_g^2 + 5 \gamma_g \gamma_d + 4 \gamma_d^2)}\,.\label{eq:m1ABasym}
    \end{align}
\end{widetext}
The known interference blockades for $m^{(1)}_A$, $m^{(1)}_{AB}$, and the leading order of $m^{(1)}_B$ arise for $\gamma_g=\gamma_d$ \cite{Synch_Drive,Synch_Entanglement}. The contributions of the coherences $\dyad{i,j}{k,l}$ to $m^{(1)}_{AB}$ are highlighted in the first line of \cref{eq:m1ABasym}. Terms originating from both $\dyad{0,1}{1,0}$ and $\dyad{-1,0}{0,-1}$ vanish individually, whereas terms proportional to $\dyad{0,0}{1,-1}$ and $\dyad{-1,1}{0,0}$ cancel in this interference blockade. For $m^{(1)}_B$, the coherences $\dyad{0,-1}{0,0}$ and $\dyad{0,0}{0,1}$ cancel with the coherences $\dyad{-1,0}{-1,1}$ and $\dyad{1,-1}{1,0}$. Note that contributions to $m^{(1)}_B$ of order $g^3 \Omega_A$ and higher do not vanish for equal gain and damping rates as we will see in the next section. There, additional coherences $\dyad{1,0}{1,1}$ and $\dyad{-1,-1}{-1,0}$ appear. The remaining terms of $m^{(1)}_B$ in the interference blockade can be interpreted as first-order synchronization $\propto\Omega_A$ of the undriven spin $B$ to the drive that acts on spin $A$ mediated via a third-order spin-spin interaction $\propto g^3$. We also recognize that the absolute values of the first moments $m^{(1)}_A$, $m^{(1)}_B$, and $m^{(1)}_{AB}$ are invariant under the exchange of the gain and damping rates.

A new coupling-induced blockade, in which contributing coherences cancel collectively, arises for certain relations between the coupling strength $g$ and the gain and damping rates. The solution of $m^{(1)}_{AB}=0$ can be obtained analytically as
\begin{align}
    \frac{\gamma_g}{\gamma_d} = \frac12(1+\sqrt{17})\frac{g^2}{\gamma_d^2} \approx \frac12(1+\sqrt{17})\frac{g^2}{(\gamma_g + \gamma_d)^2}\,, \label{eq:gblockadem1AB}
\end{align}
where the approximation holds for $g,\gamma_g\ll\gamma_d$. The solution of $m^{(1)}_{A}=0$ is obtained approximatively for (a) large $\gamma_g\gg\gamma_d$ and for (b) both large $\gamma_g,g\gg\gamma_d$
\begin{align}
    \frac{g}{\gamma_d} &\stackrel{\text{(a)}}{\approx} \frac12\sqrt{1+\sqrt{10}}, &\frac{\gamma_g}{\gamma_d}&\approx\frac{2}{\sqrt{1+\sqrt{10}}}\frac{g}{\gamma_d}\,,\label{eq:gblockadem1A1}\\
    \frac{g}{\gamma_d} &\stackrel{\text{(b)}}{\approx} 1.323, &\frac{\gamma_g}{\gamma_d} &\approx 0.7561 \frac{g}{\gamma_d}\,.\label{eq:gblockadem1A2}
\end{align}

\subsection{Equal Gain and Damping Rates}\label{sec:equalrates}
In the case of equal gain and damping rates $\gamma^{g/d}_A=\gamma^{g/d}_B=\gamma$, for $\Omega_A=0$, we obtain
\begin{align}
    \rho^\text{ss} =& \left(1 - \frac{8g^2}{8g^2+\gamma^2}\right)\rho^{(0)} + \frac{32g^2}{32g^2+4\gamma^2}\rho^{(\infty)} \nonumber\\
    &-\ii\frac{g \gamma}{16g^2+2\gamma^2}[S^+_A S^-_B + \mathrm{H.c.},\rho^{(0)}]\,,
\end{align}
where
\begin{align}
    \rho^{(\infty)} =& \frac18 \sum_{J=1,2}\sum_{M=-1,1} \ket{J,M}_c\bra{J,M}_c \nonumber\\
    &+ \frac14\sum_{J=0,2}\ket{J,0}_c\bra{J,0}_c = \frac{1}{32}(S^+_A S^-_B + S^-_A S^+_B)^2\label{eq:rhoinfg}
\end{align}
is the state in the limit $g\gg\gamma$ and is diagonal in the combined spin basis $\ket{J,M}_c$ of two spins 1. We now add a drive that acts on spin $A$ as a small perturbation. This results in the following leading-order contributions in $\Omega_A/\gamma$ to the moments of the synchronization measure of the undriven spin $B$ 
\begin{align}
    m^{(1)}_B &\approx \frac{3}{4}\frac{g^3 \Omega_A (64 g^4 + 348 g^2 \gamma^2 + 135 \gamma^4)}{(8 g^2 + \gamma^2)(4 g^2 + 9 \gamma^2)(16 g^4 + 72 g^2 \gamma^2 + 9 \gamma^4)}\nonumber\\
    &\stackrel{g\gg\gamma}{\approx} \frac{3\Omega_A}{32g} \nonumber\\
    &\stackrel{g\ll\gamma}{\approx} \frac{5 g^3 \Omega_A}{4\gamma^4}\,,\label{eq:m1Bequal}\\
    m^{(2)}_B &\approx \frac{3}{2\pi}\frac{g^2\Omega_A^2}{(g^2 + \gamma^2)(4 g^2 + \gamma^2)}\nonumber\\
    &~~\times \frac{96 g^8 + 656 g^6 \gamma^2 + 518 g^4\gamma^4 + 108 g^2\gamma^6 + 81 \gamma^8}{(8 g^2 + \gamma^2)(4 g^2 + 9 \gamma^2)(16 g^4 + 72 g^2 \gamma^2 + 9 \gamma^4)}\nonumber\\
    &\stackrel{g\gg\gamma}{\approx} \frac{9\Omega_A^2}{128\pi g^2} \nonumber\\
    &\stackrel{g\ll\gamma}{\approx} \frac{3g^2 \Omega_A^2}{2\pi\gamma^4}\,.
\end{align}
The undriven spin $B$ exhibits a 1:1 phase locking to the drive with a magnitude that to leading order is linear in $\Omega_A/\gamma$. The second moment of the combined synchronization measure for both spins is up to second order in $\Omega_A/\gamma$ proportional to
\begin{align}
    m^{(2)}_{AB} &\approx \frac{1}{8\pi}\frac{g^2}{8g^2+\gamma^2}\nonumber\\
    \times&\left(1-\frac{\Omega_A^2 (848 g^6 + 4600 g^4 \gamma^2 + 1905 g^2\gamma^4 + 702\gamma^6)}{(8 g^2 + \gamma^2)(4 g^2 + 9 \gamma^2)(16 g^4 + 72 g^2 \gamma^2 + 9 \gamma^4)}\right)\nonumber\\
    &\stackrel{g\gg\gamma}{\approx} \frac{1}{64\pi} - \frac{53\Omega_A^2 + 4\gamma^2}{2048\pi g^2}\nonumber\\
    &\stackrel{g\ll\gamma}{\approx} \frac{g^2}{8\pi\gamma^2}\left(1 - \frac{26\Omega_A^2+24 g^2}{3\gamma^4}\right)\,.
\end{align}
Analogously, for a vanishing spin-spin interaction strength $g=0$, we obtain
\begin{align}
    \rho^\text{ss} =& \left(1 - \frac{8\Omega_A^2}{8\Omega_A^2+\gamma^2}\right)\rho^{(0)} - \frac{\Omega_A^2}{8\Omega_A^2+\gamma^2}[S^+_A + S^-_A,\rho^{(0)}]^2 \nonumber\\
    &-\ii\frac{\Omega_A \gamma}{8\Omega_A^2+\gamma^2}[S^+_A + S^-_A,\rho^{(0)}]\,,
\end{align}
leading to the following contribution to the second moment of the synchronization measure of the driven spin $A$ up to second order in $g/\gamma$,
\begin{align}
    m^{(2)}_A &\approx \frac{1}{2\pi}\frac{\Omega_A^2}{8\Omega_A^2+\gamma^2}\nonumber\\
    &~~\times\left(1-\frac{g^2 (448 \Omega_A^4 + 456 \Omega_A^2 \gamma^2 + 189\gamma^2)}{(8 \Omega_A^2 + \gamma^2)(16 \Omega_A^4 + 30 \Omega_A^2 \gamma^2 + 9 \gamma^4)}\right)\nonumber\\
    &\stackrel{\Omega_A\gg\gamma}{\approx} \frac{1}{16\pi} - \frac{28g^2 + \gamma^2}{128\pi\Omega_A^2}\nonumber\\
    &\stackrel{\Omega_A\ll\gamma}{\approx} \frac{\Omega_A^2}{2\pi\gamma^2}\left(1 - \frac{21 g^2+8\Omega_A^2}{\gamma^4}\right)\,.
\end{align}
Note that for equal gain and damping rates, the first moment $m^{(1)}_A$ of the synchronization measure of the single spin $A$ and $m^{(1)}_{AB}$ of the combined synchronization measure vanish, i.e., the system is in the quantum interference blockade.

\begin{figure}[t]
    \includegraphics[width=8.6cm]{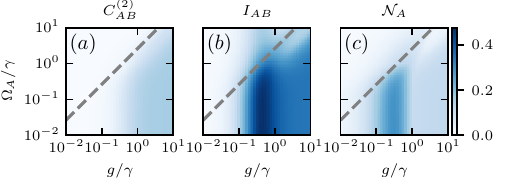}
    \caption{(a) Correlations $C^{(2)}_{AB}$ related to \crefs{fig:S_moments_two_spins}(d) to (f). (b) Quantum mutual information of spin $A$ and $B$. (c) Negativity of spin $A$. All measures are evaluated for the steady state of the Lindblad master equation. The gray dashed line denotes the theoretical prediction $\Omega_A=5\pi g/6$ of $|m^{(2)}_B/m^{(1)}_B|=1$, cf.~\cref{fig:S_moments_two_spins}.}
    \label{fig:ent_two_spin1}
\end{figure}
\begin{figure}[t]
    \includegraphics[width=8.6cm]{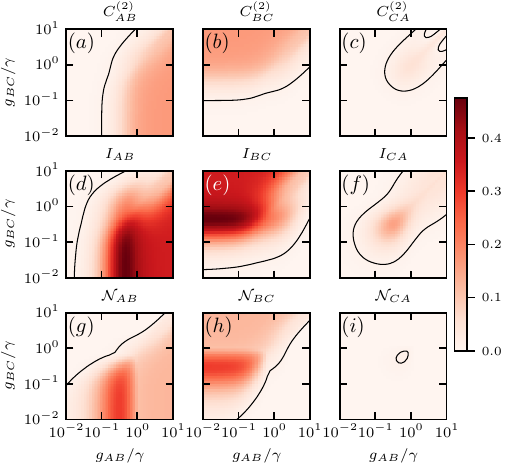}
    \caption{(a)--(c) Correlations related to \crefs{fig:S_moments_three_spins}(d) to (f). (d)--(f) Quantum mutual information of pairs of spins. (g)--(i) Negativity of pairs of spins. The black curves are contour lines at $0.01$.}
    \label{fig:ent_three_spin1}
\end{figure}

\section{Entanglement Measures}\label{sec:entanglement}
In this appendix, we compute correlations
\begin{align}
    C^{(n)}_{ij} &= \frac{\text{COV}^{(n)}_{ij}}{\sqrt{\text{COV}^{(n)}_{ii}\text{COV}^{(n)}_{jj}}}\,,\\
    \text{COV}^{(n)}_{ij} &= \langle (S^-_i S^+_j)^n \rangle - \langle (S^-_i)^n \rangle\langle (S^+_j)^n \rangle\,,
\end{align}
and entanglement measures 
\begin{align}
    I_{ij} &= S(\rho_i:\rho_j) = S(\rho_i) + S(\rho_j) - S(\rho_{ij})\,,\\
    \mathcal{N}_j(\rho) &= \frac{||\rho^{T_j}||_1 - 1}{2} = \sum_k \frac{|\lambda_k| - \lambda_k}{2}\,,
\end{align}
where $I_{ij}$ is the quantum mutual information, $S(\rho)$ is the von Neumann entropy, and $\mathcal{N}_j$ is the negativity. The eigenvalues of $\rho^{T_j}$ are denoted by $\lambda_k$, where $T_j$ indicates the partial transpose that only acts on subsystem $j$. Note that in a two-partite system, $\rho^{T_A}$ and $\rho^{T_B}=(\rho^{T_A})^T$ have the same eigenvalues and therefore $\mathcal{N}_A=\mathcal{N}_B$. For quantum systems of dimensions larger than $2\times 3$, a necessary condition of separability is zero negativity \cite{PhysRevLett.77.1413,HORODECKI19961}. Therefore, $\mathcal{N}_j>0$ implies entanglement. For mixed states, both entanglement and classical correlations contribute to the quantum mutual information $I_{ij}$.

We want to highlight the following features of correlations between both spins in the two-spin setup. In \cref{fig:ent_two_spin1}, both $I_{AB}$ and $\mathcal{N}_A$ exhibit a local maximum between $0.1\gamma<g<\gamma$ and below the gray dashed line that indicates the theoretical prediction $\Omega_A = 5\pi g/6$ of $|m^{(2)}_B/m^{(1)}_B|=1$. In this region, the first moment $m^{(1)}_B$ of the synchronization measure of spin $B$, indicating 1:1 phase locking, dominates and $p_\text{max}(\rho^{(0)})$ exhibits a strong change. Comparing all three panels of \cref{fig:ent_two_spin1}, in this system, the mutual information $I_{AB}$ appears to be a combination of correlations, e.g., $C^{(2)}_{AB}$, and entanglement.

In \cref{fig:ent_three_spin1}, we present the correlations, quantum mutual information, and negativity between pairs of spin-1 oscillators. We define $\mathcal{N}_{ij}$ as the negativity of spin $i$ evaluated for the reduced density matrix of the subsystem of spin $i$ and $j$. The correlations, mutual information, and negativity of subsystem $AB$ ($BC$) exhibit similar qualitative features, e.g., a local maximum between $0.1\gamma<g_{AB}~(g_{BC})<\gamma$, as in the two-spin case. The measures of subsystem $CA$ exhibit local maxima at $0.1\gamma<g_{AB},g_{BC}<\gamma$. Here, qualitatively, the measures of the other two subsystems overlap.


\begin{thebibliography}{36}%
	\makeatletter
	\providecommand \@ifxundefined [1]{%
		\@ifx{#1\undefined}
	}%
	\providecommand \@ifnum [1]{%
		\ifnum #1\expandafter \@firstoftwo
		\else \expandafter \@secondoftwo
		\fi
	}%
	\providecommand \@ifx [1]{%
		\ifx #1\expandafter \@firstoftwo
		\else \expandafter \@secondoftwo
		\fi
	}%
	\providecommand \natexlab [1]{#1}%
	\providecommand \enquote  [1]{``#1''}%
	\providecommand \bibnamefont  [1]{#1}%
	\providecommand \bibfnamefont [1]{#1}%
	\providecommand \citenamefont [1]{#1}%
	\providecommand \href@noop [0]{\@secondoftwo}%
	\providecommand \href [0]{\begingroup \@sanitize@url \@href}%
	\providecommand \@href[1]{\@@startlink{#1}\@@href}%
	\providecommand \@@href[1]{\endgroup#1\@@endlink}%
	\providecommand \@sanitize@url [0]{\catcode `\\12\catcode `\$12\catcode
		`\&12\catcode `\#12\catcode `\^12\catcode `\_12\catcode `\%12\relax}%
	\providecommand \@@startlink[1]{}%
	\providecommand \@@endlink[0]{}%
	\providecommand \url  [0]{\begingroup\@sanitize@url \@url }%
	\providecommand \@url [1]{\endgroup\@href {#1}{\urlprefix }}%
	\providecommand \urlprefix  [0]{URL }%
	\providecommand \Eprint [0]{\href }%
	\providecommand \doibase [0]{https://doi.org/}%
	\providecommand \selectlanguage [0]{\@gobble}%
	\providecommand \bibinfo  [0]{\@secondoftwo}%
	\providecommand \bibfield  [0]{\@secondoftwo}%
	\providecommand \translation [1]{[#1]}%
	\providecommand \BibitemOpen [0]{}%
	\providecommand \bibitemStop [0]{}%
	\providecommand \bibitemNoStop [0]{.\EOS\space}%
	\providecommand \EOS [0]{\spacefactor3000\relax}%
	\providecommand \BibitemShut  [1]{\csname bibitem#1\endcsname}%
	\let\auto@bib@innerbib\@empty
	\bibitem [{\citenamefont {Goldsztein}\ \emph {et~al.}(2021)\citenamefont
		{Goldsztein}, \citenamefont {Nadeau},\ and\ \citenamefont
		{Strogatz}}]{10.1063/5.0026335}%
	\BibitemOpen
	\bibfield  {author} {\bibinfo {author} {\bibfnamefont {G.~H.}\ \bibnamefont
			{Goldsztein}}, \bibinfo {author} {\bibfnamefont {A.~N.}\ \bibnamefont
			{Nadeau}},\ and\ \bibinfo {author} {\bibfnamefont {S.~H.}\ \bibnamefont
			{Strogatz}},\ }\bibfield  {title} {\bibinfo {title} {{Synchronization of
				clocks and metronomes: A perturbation analysis based on multiple
				timescales}},\ }\href {https://doi.org/10.1063/5.0026335} {\bibfield
		{journal} {\bibinfo  {journal} {Chaos}\ }\textbf {\bibinfo {volume} {31}},\
		\bibinfo {pages} {023109} (\bibinfo {year} {2021})}\BibitemShut {NoStop}%
	\bibitem [{\citenamefont {Buck}(1938)}]{10.1086/394562}%
	\BibitemOpen
	\bibfield  {author} {\bibinfo {author} {\bibfnamefont {J.~B.}\ \bibnamefont
			{Buck}},\ }\bibfield  {title} {\bibinfo {title} {Synchronous rhythmic
			flashing of fireflies},\ }\href {https://doi.org/10.1086/394562} {\bibfield
		{journal} {\bibinfo  {journal} {The Quarterly Review of Biology}\ }\textbf
		{\bibinfo {volume} {13}},\ \bibinfo {pages} {301} (\bibinfo {year}
		{1938})}\BibitemShut {NoStop}%
	\bibitem [{\citenamefont {{Strogatz}}\ \emph {et~al.}(2005)\citenamefont
		{{Strogatz}}, \citenamefont {{Abrams}}, \citenamefont {{McRobie}},
		\citenamefont {{Eckhardt}},\ and\ \citenamefont
		{{Ott}}}]{2005Natur.438...43S}%
	\BibitemOpen
	\bibfield  {author} {\bibinfo {author} {\bibfnamefont {S.~H.}\ \bibnamefont
			{{Strogatz}}}, \bibinfo {author} {\bibfnamefont {D.~M.}\ \bibnamefont
			{{Abrams}}}, \bibinfo {author} {\bibfnamefont {A.}~\bibnamefont {{McRobie}}},
		\bibinfo {author} {\bibfnamefont {B.}~\bibnamefont {{Eckhardt}}},\ and\
		\bibinfo {author} {\bibfnamefont {E.}~\bibnamefont {{Ott}}},\ }\bibfield
	{title} {\bibinfo {title} {{Theoretical mechanics: Crowd synchrony on the
				Millennium Bridge}},\ }\href {https://doi.org/10.1038/43843a} {\bibfield
		{journal} {\bibinfo  {journal} {\nat}\ }\textbf {\bibinfo {volume} {438}},\
		\bibinfo {pages} {43} (\bibinfo {year} {2005})}\BibitemShut {NoStop}%
	\bibitem [{\citenamefont {Wang}(2022)}]{CellCycleSynch}%
	\BibitemOpen
	\bibinfo {editor} {\bibfnamefont {Z.}~\bibnamefont {Wang}},\ ed.,\ \href
	{https://doi.org/https://doi.org/10.1007/978-1-0716-2736-5} {\emph {\bibinfo
			{title} {Cell-Cycle Synchronization}}},\ Methods in Molecular Biology\
	(\bibinfo  {publisher} {Humana},\ \bibinfo {address} {New York},\ \bibinfo
	{year} {2022})\BibitemShut {NoStop}%
	\bibitem [{\citenamefont {Acebr\'on}\ \emph {et~al.}(2005)\citenamefont
		{Acebr\'on}, \citenamefont {Bonilla}, \citenamefont {P\'erez~Vicente},
		\citenamefont {Ritort},\ and\ \citenamefont {Spigler}}]{RevModPhys.77.137}%
	\BibitemOpen
	\bibfield  {author} {\bibinfo {author} {\bibfnamefont {J.~A.}\ \bibnamefont
			{Acebr\'on}}, \bibinfo {author} {\bibfnamefont {L.~L.}\ \bibnamefont
			{Bonilla}}, \bibinfo {author} {\bibfnamefont {C.~J.}\ \bibnamefont
			{P\'erez~Vicente}}, \bibinfo {author} {\bibfnamefont {F.}~\bibnamefont
			{Ritort}},\ and\ \bibinfo {author} {\bibfnamefont {R.}~\bibnamefont
			{Spigler}},\ }\bibfield  {title} {\bibinfo {title} {The {Kuramoto} model: A
			simple paradigm for synchronization phenomena},\ }\href
	{https://doi.org/10.1103/RevModPhys.77.137} {\bibfield  {journal} {\bibinfo
			{journal} {Rev. Mod. Phys.}\ }\textbf {\bibinfo {volume} {77}},\ \bibinfo
		{pages} {137} (\bibinfo {year} {2005})}\BibitemShut {NoStop}%
	\bibitem [{\citenamefont {Pikovsky}\ \emph {et~al.}(2001)\citenamefont
		{Pikovsky}, \citenamefont {Rosenblum},\ and\ \citenamefont
		{Kurths}}]{Synch_Pikovsky}%
	\BibitemOpen
	\bibfield  {author} {\bibinfo {author} {\bibfnamefont {A.}~\bibnamefont
			{Pikovsky}}, \bibinfo {author} {\bibfnamefont {M.}~\bibnamefont
			{Rosenblum}},\ and\ \bibinfo {author} {\bibfnamefont {J.}~\bibnamefont
			{Kurths}},\ }\href@noop {} {\emph {\bibinfo {title} {{Synchronization: A
					Universal Concept in Nonlinear Science}}}}\ (\bibinfo  {publisher} {Cambridge
		University Press, Cambridge, England},\ \bibinfo {year} {2001})\BibitemShut
	{NoStop}%
	\bibitem [{\citenamefont {Strogatz}(2003)}]{Synch_Strogatz}%
	\BibitemOpen
	\bibfield  {author} {\bibinfo {author} {\bibfnamefont {S.~H.}\ \bibnamefont
			{Strogatz}},\ }\href@noop {} {\emph {\bibinfo {title} {{Sync: The Emerging
					Science of Spontaneous Order}}}}\ (\bibinfo  {publisher} {Hyperion, New
		York},\ \bibinfo {year} {2003})\BibitemShut {NoStop}%
	\bibitem [{\citenamefont {Okuda}\ and\ \citenamefont
		{Kuramoto}(1991)}]{10.1143/ptp/86.6.1159}%
	\BibitemOpen
	\bibfield  {author} {\bibinfo {author} {\bibfnamefont {K.}~\bibnamefont
			{Okuda}}\ and\ \bibinfo {author} {\bibfnamefont {Y.}~\bibnamefont
			{Kuramoto}},\ }\bibfield  {title} {\bibinfo {title} {{Mutual entrainment
				between populations of coupled oscillators}},\ }\href
	{https://doi.org/10.1143/ptp/86.6.1159} {\bibfield  {journal} {\bibinfo
			{journal} {Progress of Theoretical Physics}\ }\textbf {\bibinfo {volume}
			{86}},\ \bibinfo {pages} {1159} (\bibinfo {year} {1991})}\BibitemShut
	{NoStop}%
	\bibitem [{\citenamefont {Chia}\ \emph {et~al.}(2020)\citenamefont {Chia},
		\citenamefont {Kwek},\ and\ \citenamefont {Noh}}]{PhysRevE.102.042213}%
	\BibitemOpen
	\bibfield  {author} {\bibinfo {author} {\bibfnamefont {A.}~\bibnamefont
			{Chia}}, \bibinfo {author} {\bibfnamefont {L.~C.}\ \bibnamefont {Kwek}},\
		and\ \bibinfo {author} {\bibfnamefont {C.}~\bibnamefont {Noh}},\ }\bibfield
	{title} {\bibinfo {title} {Relaxation oscillations and frequency entrainment
			in quantum mechanics},\ }\href {https://doi.org/10.1103/PhysRevE.102.042213}
	{\bibfield  {journal} {\bibinfo  {journal} {Phys. Rev. E}\ }\textbf {\bibinfo
			{volume} {102}},\ \bibinfo {pages} {042213} (\bibinfo {year}
		{2020})}\BibitemShut {NoStop}%
	\bibitem [{\citenamefont {Ben~Arosh}\ \emph {et~al.}(2021)\citenamefont
		{Ben~Arosh}, \citenamefont {Cross},\ and\ \citenamefont
		{Lifshitz}}]{PhysRevResearch.3.013130}%
	\BibitemOpen
	\bibfield  {author} {\bibinfo {author} {\bibfnamefont {L.}~\bibnamefont
			{Ben~Arosh}}, \bibinfo {author} {\bibfnamefont {M.~C.}\ \bibnamefont
			{Cross}},\ and\ \bibinfo {author} {\bibfnamefont {R.}~\bibnamefont
			{Lifshitz}},\ }\bibfield  {title} {\bibinfo {title} {Quantum limit cycles and
			the {Rayleigh} and van der {Pol} oscillators},\ }\href
	{https://doi.org/10.1103/PhysRevResearch.3.013130} {\bibfield  {journal}
		{\bibinfo  {journal} {Phys. Rev. Res.}\ }\textbf {\bibinfo {volume} {3}},\
		\bibinfo {pages} {013130} (\bibinfo {year} {2021})}\BibitemShut {NoStop}%
	\bibitem [{\citenamefont {Ludwig}\ and\ \citenamefont
		{Marquardt}(2013)}]{PhysRevLett.111.073603}%
	\BibitemOpen
	\bibfield  {author} {\bibinfo {author} {\bibfnamefont {M.}~\bibnamefont
			{Ludwig}}\ and\ \bibinfo {author} {\bibfnamefont {F.}~\bibnamefont
			{Marquardt}},\ }\bibfield  {title} {\bibinfo {title} {Quantum many-body
			dynamics in optomechanical arrays},\ }\href
	{https://doi.org/10.1103/PhysRevLett.111.073603} {\bibfield  {journal}
		{\bibinfo  {journal} {Phys. Rev. Lett.}\ }\textbf {\bibinfo {volume} {111}},\
		\bibinfo {pages} {073603} (\bibinfo {year} {2013})}\BibitemShut {NoStop}%
	\bibitem [{\citenamefont {Lee}\ and\ \citenamefont
		{Sadeghpour}(2013)}]{Synch_vdP_Lee}%
	\BibitemOpen
	\bibfield  {author} {\bibinfo {author} {\bibfnamefont {T.~E.}\ \bibnamefont
			{Lee}}\ and\ \bibinfo {author} {\bibfnamefont {H.~R.}\ \bibnamefont
			{Sadeghpour}},\ }\bibfield  {title} {\bibinfo {title} {Quantum
			synchronization of quantum van der {Pol} oscillators with trapped ions},\
	}\href {https://doi.org/10.1103/PhysRevLett.111.234101} {\bibfield  {journal}
		{\bibinfo  {journal} {Phys. Rev. Lett.}\ }\textbf {\bibinfo {volume} {111}},\
		\bibinfo {pages} {234101} (\bibinfo {year} {2013})}\BibitemShut {NoStop}%
	\bibitem [{\citenamefont {Walter}\ \emph {et~al.}(2015)\citenamefont {Walter},
		\citenamefont {Nunnenkamp},\ and\ \citenamefont {Bruder}}]{Synch_vdP_Walter}%
	\BibitemOpen
	\bibfield  {author} {\bibinfo {author} {\bibfnamefont {S.}~\bibnamefont
			{Walter}}, \bibinfo {author} {\bibfnamefont {A.}~\bibnamefont {Nunnenkamp}},\
		and\ \bibinfo {author} {\bibfnamefont {C.}~\bibnamefont {Bruder}},\
	}\bibfield  {title} {\bibinfo {title} {Quantum synchronization of two van der
			{Pol} oscillators},\ }\href
	{https://doi.org/https://doi.org/10.1002/andp.201400144} {\bibfield
		{journal} {\bibinfo  {journal} {Annalen der Physik}\ }\textbf {\bibinfo
			{volume} {527}},\ \bibinfo {pages} {131} (\bibinfo {year}
		{2015})}\BibitemShut {NoStop}%
	\bibitem [{\citenamefont {W\"achtler}\ and\ \citenamefont
		{Platero}(2023)}]{PhysRevResearch.5.023021}%
	\BibitemOpen
	\bibfield  {author} {\bibinfo {author} {\bibfnamefont {C.~W.}\ \bibnamefont
			{W\"achtler}}\ and\ \bibinfo {author} {\bibfnamefont {G.}~\bibnamefont
			{Platero}},\ }\bibfield  {title} {\bibinfo {title} {Topological
			synchronization of quantum van der pol oscillators},\ }\href
	{https://doi.org/10.1103/PhysRevResearch.5.023021} {\bibfield  {journal}
		{\bibinfo  {journal} {Phys. Rev. Res.}\ }\textbf {\bibinfo {volume} {5}},\
		\bibinfo {pages} {023021} (\bibinfo {year} {2023})}\BibitemShut {NoStop}%
	\bibitem [{\citenamefont {Roulet}\ and\ \citenamefont
		{Bruder}(2018{\natexlab{a}})}]{Synch_Drive}%
	\BibitemOpen
	\bibfield  {author} {\bibinfo {author} {\bibfnamefont {A.}~\bibnamefont
			{Roulet}}\ and\ \bibinfo {author} {\bibfnamefont {C.}~\bibnamefont
			{Bruder}},\ }\bibfield  {title} {\bibinfo {title} {Synchronizing the smallest
			possible system},\ }\href {https://doi.org/10.1103/PhysRevLett.121.053601}
	{\bibfield  {journal} {\bibinfo  {journal} {Phys. Rev. Lett.}\ }\textbf
		{\bibinfo {volume} {121}},\ \bibinfo {pages} {053601} (\bibinfo {year}
		{2018}{\natexlab{a}})}\BibitemShut {NoStop}%
	\bibitem [{\citenamefont {Roulet}\ and\ \citenamefont
		{Bruder}(2018{\natexlab{b}})}]{Synch_Entanglement}%
	\BibitemOpen
	\bibfield  {author} {\bibinfo {author} {\bibfnamefont {A.}~\bibnamefont
			{Roulet}}\ and\ \bibinfo {author} {\bibfnamefont {C.}~\bibnamefont
			{Bruder}},\ }\bibfield  {title} {\bibinfo {title} {Quantum synchronization
			and entanglement generation},\ }\href
	{https://doi.org/10.1103/PhysRevLett.121.063601} {\bibfield  {journal}
		{\bibinfo  {journal} {Phys. Rev. Lett.}\ }\textbf {\bibinfo {volume} {121}},\
		\bibinfo {pages} {063601} (\bibinfo {year} {2018}{\natexlab{b}})}\BibitemShut
	{NoStop}%
	\bibitem [{\citenamefont {Parra-L\'opez}\ and\ \citenamefont
		{Bergli}(2020)}]{PhysRevA.101.062104}%
	\BibitemOpen
	\bibfield  {author} {\bibinfo {author} {\bibfnamefont {A.}~\bibnamefont
			{Parra-L\'opez}}\ and\ \bibinfo {author} {\bibfnamefont {J.}~\bibnamefont
			{Bergli}},\ }\bibfield  {title} {\bibinfo {title} {Synchronization in
			two-level quantum systems},\ }\href
	{https://doi.org/10.1103/PhysRevA.101.062104} {\bibfield  {journal} {\bibinfo
			{journal} {Phys. Rev. A}\ }\textbf {\bibinfo {volume} {101}},\ \bibinfo
		{pages} {062104} (\bibinfo {year} {2020})}\BibitemShut {NoStop}%
	\bibitem [{\citenamefont {Laskar}\ \emph {et~al.}(2020)\citenamefont {Laskar},
		\citenamefont {Adhikary}, \citenamefont {Mondal}, \citenamefont {Katiyar},
		\citenamefont {Vinjanampathy},\ and\ \citenamefont
		{Ghosh}}]{PhysRevLett.125.013601}%
	\BibitemOpen
	\bibfield  {author} {\bibinfo {author} {\bibfnamefont {A.~W.}\ \bibnamefont
			{Laskar}}, \bibinfo {author} {\bibfnamefont {P.}~\bibnamefont {Adhikary}},
		\bibinfo {author} {\bibfnamefont {S.}~\bibnamefont {Mondal}}, \bibinfo
		{author} {\bibfnamefont {P.}~\bibnamefont {Katiyar}}, \bibinfo {author}
		{\bibfnamefont {S.}~\bibnamefont {Vinjanampathy}},\ and\ \bibinfo {author}
		{\bibfnamefont {S.}~\bibnamefont {Ghosh}},\ }\bibfield  {title} {\bibinfo
		{title} {Observation of quantum phase synchronization in spin-1 atoms},\
	}\href {https://doi.org/10.1103/PhysRevLett.125.013601} {\bibfield  {journal}
		{\bibinfo  {journal} {Phys. Rev. Lett.}\ }\textbf {\bibinfo {volume} {125}},\
		\bibinfo {pages} {013601} (\bibinfo {year} {2020})}\BibitemShut {NoStop}%
	\bibitem [{\citenamefont {Krithika}\ \emph {et~al.}(2022)\citenamefont
		{Krithika}, \citenamefont {Solanki}, \citenamefont {Vinjanampathy},\ and\
		\citenamefont {Mahesh}}]{PhysRevA.105.062206}%
	\BibitemOpen
	\bibfield  {author} {\bibinfo {author} {\bibfnamefont {V.~R.}\ \bibnamefont
			{Krithika}}, \bibinfo {author} {\bibfnamefont {P.}~\bibnamefont {Solanki}},
		\bibinfo {author} {\bibfnamefont {S.}~\bibnamefont {Vinjanampathy}},\ and\
		\bibinfo {author} {\bibfnamefont {T.~S.}\ \bibnamefont {Mahesh}},\ }\bibfield
	{title} {\bibinfo {title} {Observation of quantum phase synchronization in a
			nuclear-spin system},\ }\href {https://doi.org/10.1103/PhysRevA.105.062206}
	{\bibfield  {journal} {\bibinfo  {journal} {Phys. Rev. A}\ }\textbf {\bibinfo
			{volume} {105}},\ \bibinfo {pages} {062206} (\bibinfo {year}
		{2022})}\BibitemShut {NoStop}%
	\bibitem [{\citenamefont {Zhang}\ \emph {et~al.}(2023)\citenamefont {Zhang},
		\citenamefont {Wang}, \citenamefont {Wang}, \citenamefont {Zhang},
		\citenamefont {Wu}, \citenamefont {Jie},\ and\ \citenamefont
		{Lu}}]{PhysRevResearch.5.033209}%
	\BibitemOpen
	\bibfield  {author} {\bibinfo {author} {\bibfnamefont {L.}~\bibnamefont
			{Zhang}}, \bibinfo {author} {\bibfnamefont {Z.}~\bibnamefont {Wang}},
		\bibinfo {author} {\bibfnamefont {Y.}~\bibnamefont {Wang}}, \bibinfo {author}
		{\bibfnamefont {J.}~\bibnamefont {Zhang}}, \bibinfo {author} {\bibfnamefont
			{Z.}~\bibnamefont {Wu}}, \bibinfo {author} {\bibfnamefont {J.}~\bibnamefont
			{Jie}},\ and\ \bibinfo {author} {\bibfnamefont {Y.}~\bibnamefont {Lu}},\
	}\bibfield  {title} {\bibinfo {title} {Quantum synchronization of a single
			trapped-ion qubit},\ }\href
	{https://doi.org/10.1103/PhysRevResearch.5.033209} {\bibfield  {journal}
		{\bibinfo  {journal} {Phys. Rev. Res.}\ }\textbf {\bibinfo {volume} {5}},\
		\bibinfo {pages} {033209} (\bibinfo {year} {2023})}\BibitemShut {NoStop}%
	\bibitem [{\citenamefont {Koppenh\"ofer}\ \emph {et~al.}(2020)\citenamefont
		{Koppenh\"ofer}, \citenamefont {Bruder},\ and\ \citenamefont
		{Roulet}}]{PhysRevResearch.2.023026}%
	\BibitemOpen
	\bibfield  {author} {\bibinfo {author} {\bibfnamefont {M.}~\bibnamefont
			{Koppenh\"ofer}}, \bibinfo {author} {\bibfnamefont {C.}~\bibnamefont
			{Bruder}},\ and\ \bibinfo {author} {\bibfnamefont {A.}~\bibnamefont
			{Roulet}},\ }\bibfield  {title} {\bibinfo {title} {Quantum synchronization on
			the {IBM} {Q} system},\ }\href
	{https://doi.org/10.1103/PhysRevResearch.2.023026} {\bibfield  {journal}
		{\bibinfo  {journal} {Phys. Rev. Res.}\ }\textbf {\bibinfo {volume} {2}},\
		\bibinfo {pages} {023026} (\bibinfo {year} {2020})}\BibitemShut {NoStop}%
	\bibitem [{\citenamefont {Koppenh\"ofer}\ and\ \citenamefont
		{Roulet}(2019)}]{PhysRevA.99.043804}%
	\BibitemOpen
	\bibfield  {author} {\bibinfo {author} {\bibfnamefont {M.}~\bibnamefont
			{Koppenh\"ofer}}\ and\ \bibinfo {author} {\bibfnamefont {A.}~\bibnamefont
			{Roulet}},\ }\bibfield  {title} {\bibinfo {title} {Optimal synchronization
			deep in the quantum regime: Resource and fundamental limit},\ }\href
	{https://doi.org/10.1103/PhysRevA.99.043804} {\bibfield  {journal} {\bibinfo
			{journal} {Phys. Rev. A}\ }\textbf {\bibinfo {volume} {99}},\ \bibinfo
		{pages} {043804} (\bibinfo {year} {2019})}\BibitemShut {NoStop}%
	\bibitem [{\citenamefont {Solanki}\ \emph {et~al.}(2023)\citenamefont
		{Solanki}, \citenamefont {Mehdi}, \citenamefont {Hajdu\ifmmode~\check{s}\else
			\v{s}\fi{}ek},\ and\ \citenamefont {Vinjanampathy}}]{PhysRevA.108.022216}%
	\BibitemOpen
	\bibfield  {author} {\bibinfo {author} {\bibfnamefont {P.}~\bibnamefont
			{Solanki}}, \bibinfo {author} {\bibfnamefont {F.~M.}\ \bibnamefont {Mehdi}},
		\bibinfo {author} {\bibfnamefont {M.}~\bibnamefont
			{Hajdu\ifmmode~\check{s}\else \v{s}\fi{}ek}},\ and\ \bibinfo {author}
		{\bibfnamefont {S.}~\bibnamefont {Vinjanampathy}},\ }\bibfield  {title}
	{\bibinfo {title} {Symmetries and synchronization blockade},\ }\href
	{https://doi.org/10.1103/PhysRevA.108.022216} {\bibfield  {journal} {\bibinfo
			{journal} {Phys. Rev. A}\ }\textbf {\bibinfo {volume} {108}},\ \bibinfo
		{pages} {022216} (\bibinfo {year} {2023})}\BibitemShut {NoStop}%
	\bibitem [{\citenamefont {L\"orch}\ \emph {et~al.}(2017)\citenamefont
		{L\"orch}, \citenamefont {Nigg}, \citenamefont {Nunnenkamp}, \citenamefont
		{Tiwari},\ and\ \citenamefont {Bruder}}]{PhysRevLett.118.243602}%
	\BibitemOpen
	\bibfield  {author} {\bibinfo {author} {\bibfnamefont {N.}~\bibnamefont
			{L\"orch}}, \bibinfo {author} {\bibfnamefont {S.~E.}\ \bibnamefont {Nigg}},
		\bibinfo {author} {\bibfnamefont {A.}~\bibnamefont {Nunnenkamp}}, \bibinfo
		{author} {\bibfnamefont {R.~P.}\ \bibnamefont {Tiwari}},\ and\ \bibinfo
		{author} {\bibfnamefont {C.}~\bibnamefont {Bruder}},\ }\bibfield  {title}
	{\bibinfo {title} {Quantum synchronization blockade: Energy quantization
			hinders synchronization of identical oscillators},\ }\href
	{https://doi.org/10.1103/PhysRevLett.118.243602} {\bibfield  {journal}
		{\bibinfo  {journal} {Phys. Rev. Lett.}\ }\textbf {\bibinfo {volume} {118}},\
		\bibinfo {pages} {243602} (\bibinfo {year} {2017})}\BibitemShut {NoStop}%
	\bibitem [{\citenamefont {Weiss}\ \emph {et~al.}(2016)\citenamefont {Weiss},
		\citenamefont {Kronwald},\ and\ \citenamefont {Marquardt}}]{Weiss_2016}%
	\BibitemOpen
	\bibfield  {author} {\bibinfo {author} {\bibfnamefont {T.}~\bibnamefont
			{Weiss}}, \bibinfo {author} {\bibfnamefont {A.}~\bibnamefont {Kronwald}},\
		and\ \bibinfo {author} {\bibfnamefont {F.}~\bibnamefont {Marquardt}},\
	}\bibfield  {title} {\bibinfo {title} {Noise-induced transitions in
			optomechanical synchronization},\ }\href
	{https://doi.org/10.1088/1367-2630/18/1/013043} {\bibfield  {journal}
		{\bibinfo  {journal} {New Journal of Physics}\ }\textbf {\bibinfo {volume}
			{18}},\ \bibinfo {pages} {013043} (\bibinfo {year} {2016})}\BibitemShut
	{NoStop}%
	\bibitem [{\citenamefont {Hush}\ \emph {et~al.}(2015)\citenamefont {Hush},
		\citenamefont {Li}, \citenamefont {Genway}, \citenamefont {Lesanovsky},\ and\
		\citenamefont {Armour}}]{phase_dist_Hush}%
	\BibitemOpen
	\bibfield  {author} {\bibinfo {author} {\bibfnamefont {M.~R.}\ \bibnamefont
			{Hush}}, \bibinfo {author} {\bibfnamefont {W.}~\bibnamefont {Li}}, \bibinfo
		{author} {\bibfnamefont {S.}~\bibnamefont {Genway}}, \bibinfo {author}
		{\bibfnamefont {I.}~\bibnamefont {Lesanovsky}},\ and\ \bibinfo {author}
		{\bibfnamefont {A.~D.}\ \bibnamefont {Armour}},\ }\bibfield  {title}
	{\bibinfo {title} {Spin correlations as a probe of quantum synchronization in
			trapped-ion phonon lasers},\ }\href
	{https://doi.org/10.1103/PhysRevA.91.061401} {\bibfield  {journal} {\bibinfo
			{journal} {Phys. Rev. A}\ }\textbf {\bibinfo {volume} {91}},\ \bibinfo
		{pages} {061401(R)} (\bibinfo {year} {2015})}\BibitemShut {NoStop}%
	\bibitem [{\citenamefont {Chepelianskii}\ and\ \citenamefont
		{Shepelyansky}(2024)}]{Chepelianskii2024}%
	\BibitemOpen
	\bibfield  {author} {\bibinfo {author} {\bibfnamefont {A.~D.}\ \bibnamefont
			{Chepelianskii}}\ and\ \bibinfo {author} {\bibfnamefont {D.~L.}\ \bibnamefont
			{Shepelyansky}},\ }\bibfield  {title} {\bibinfo {title} {Quantum
			synchronization and entanglement of dissipative qubits coupled to a
			resonator},\ }\href {https://doi.org/10.3390/e26050415} {\bibfield  {journal}
		{\bibinfo  {journal} {Entropy}\ }\textbf {\bibinfo {volume} {26}},\ \bibinfo
		{pages} {415} (\bibinfo {year} {2024})}\BibitemShut {NoStop}%
	\bibitem [{\citenamefont {Mari}\ \emph {et~al.}(2013)\citenamefont {Mari},
		\citenamefont {Farace}, \citenamefont {Didier}, \citenamefont {Giovannetti},\
		and\ \citenamefont {Fazio}}]{PhysRevLett.111.103605}%
	\BibitemOpen
	\bibfield  {author} {\bibinfo {author} {\bibfnamefont {A.}~\bibnamefont
			{Mari}}, \bibinfo {author} {\bibfnamefont {A.}~\bibnamefont {Farace}},
		\bibinfo {author} {\bibfnamefont {N.}~\bibnamefont {Didier}}, \bibinfo
		{author} {\bibfnamefont {V.}~\bibnamefont {Giovannetti}},\ and\ \bibinfo
		{author} {\bibfnamefont {R.}~\bibnamefont {Fazio}},\ }\bibfield  {title}
	{\bibinfo {title} {Measures of quantum synchronization in continuous variable
			systems},\ }\href {https://doi.org/10.1103/PhysRevLett.111.103605} {\bibfield
		{journal} {\bibinfo  {journal} {Phys. Rev. Lett.}\ }\textbf {\bibinfo
			{volume} {111}},\ \bibinfo {pages} {103605} (\bibinfo {year}
		{2013})}\BibitemShut {NoStop}%
	\bibitem [{\citenamefont {Lee}\ \emph {et~al.}(2014)\citenamefont {Lee},
		\citenamefont {Chan},\ and\ \citenamefont {Wang}}]{PhysRevE.89.022913}%
	\BibitemOpen
	\bibfield  {author} {\bibinfo {author} {\bibfnamefont {T.~E.}\ \bibnamefont
			{Lee}}, \bibinfo {author} {\bibfnamefont {C.-K.}\ \bibnamefont {Chan}},\ and\
		\bibinfo {author} {\bibfnamefont {S.}~\bibnamefont {Wang}},\ }\bibfield
	{title} {\bibinfo {title} {Entanglement tongue and quantum synchronization of
			disordered oscillators},\ }\href {https://doi.org/10.1103/PhysRevE.89.022913}
	{\bibfield  {journal} {\bibinfo  {journal} {Phys. Rev. E}\ }\textbf {\bibinfo
			{volume} {89}},\ \bibinfo {pages} {022913} (\bibinfo {year}
		{2014})}\BibitemShut {NoStop}%
	\bibitem [{\citenamefont {Garg}\ \emph {et~al.}(2023)\citenamefont {Garg},
		\citenamefont {Manju}, \citenamefont {Dasgupta},\ and\ \citenamefont
		{Biswas}}]{GARG2023128557}%
	\BibitemOpen
	\bibfield  {author} {\bibinfo {author} {\bibfnamefont {D.}~\bibnamefont
			{Garg}}, \bibinfo {author} {\bibnamefont {Manju}}, \bibinfo {author}
		{\bibfnamefont {S.}~\bibnamefont {Dasgupta}},\ and\ \bibinfo {author}
		{\bibfnamefont {A.}~\bibnamefont {Biswas}},\ }\bibfield  {title} {\bibinfo
		{title} {Quantum synchronization and entanglement of indirectly coupled
			mechanical oscillators in cavity optomechanics: A numerical study},\ }\href
	{https://doi.org/https://doi.org/10.1016/j.physleta.2022.128557} {\bibfield
		{journal} {\bibinfo  {journal} {Physics Letters A}\ }\textbf {\bibinfo
			{volume} {457}},\ \bibinfo {pages} {128557} (\bibinfo {year}
		{2023})}\BibitemShut {NoStop}%
	\bibitem [{\citenamefont {Johansson}\ \emph {et~al.}(2013)\citenamefont
		{Johansson}, \citenamefont {Nation},\ and\ \citenamefont {Nori}}]{QuTiP}%
	\BibitemOpen
	\bibfield  {author} {\bibinfo {author} {\bibfnamefont {J.}~\bibnamefont
			{Johansson}}, \bibinfo {author} {\bibfnamefont {P.}~\bibnamefont {Nation}},\
		and\ \bibinfo {author} {\bibfnamefont {F.}~\bibnamefont {Nori}},\ }\bibfield
	{title} {\bibinfo {title} {{QuTiP} 2: A {Python} framework for the dynamics
			of open quantum systems},\ }\href
	{https://doi.org/https://doi.org/10.1016/j.cpc.2012.11.019} {\bibfield
		{journal} {\bibinfo  {journal} {Computer Physics Communications}\ }\textbf
		{\bibinfo {volume} {184}},\ \bibinfo {pages} {1234} (\bibinfo {year}
		{2013})}\BibitemShut {NoStop}%
	\bibitem [{\citenamefont {Wigner}(1959)}]{Wigner1959}%
	\BibitemOpen
	\bibfield  {author} {\bibinfo {author} {\bibfnamefont {E.~P.}\ \bibnamefont
			{Wigner}},\ }\href@noop {} {\emph {\bibinfo {title} {{Group theory: And its
					application to the quantum mechanics of atomic spectra}}}}\ (\bibinfo
	{publisher} {Academic Press},\ \bibinfo {address} {New York},\ \bibinfo
	{year} {1959})\BibitemShut {NoStop}%
	\bibitem [{\citenamefont {Gradshteyn}\ and\ \citenamefont
		{Ryzhik}(2015)}]{GradshteynRyzhik}%
	\BibitemOpen
	\bibfield  {author} {\bibinfo {author} {\bibfnamefont {I.}~\bibnamefont
			{Gradshteyn}}\ and\ \bibinfo {author} {\bibfnamefont {I.}~\bibnamefont
			{Ryzhik}},\ }\href {https://doi.org/https://doi.org/10.1016/C2010-0-64839-5}
	{\emph {\bibinfo {title} {Table of Integrals, Series, and Products (Eighth
				Edition)}}},\ edited by\ \bibinfo {editor} {\bibfnamefont {D.}~\bibnamefont
		{Zwillinger}}\ and\ \bibinfo {editor} {\bibfnamefont {V.}~\bibnamefont
		{Moll}}\ (\bibinfo  {publisher} {Academic Press},\ \bibinfo {address}
	{Boston},\ \bibinfo {year} {2015})\BibitemShut {NoStop}%
	\bibitem [{\citenamefont {Tan}\ \emph {et~al.}(2022)\citenamefont {Tan},
		\citenamefont {Bruder},\ and\ \citenamefont
		{Koppenh{\"{o}}fer}}]{Tan2022halfintegervs}%
	\BibitemOpen
	\bibfield  {author} {\bibinfo {author} {\bibfnamefont {R.}~\bibnamefont
			{Tan}}, \bibinfo {author} {\bibfnamefont {C.}~\bibnamefont {Bruder}},\ and\
		\bibinfo {author} {\bibfnamefont {M.}~\bibnamefont {Koppenh{\"{o}}fer}},\
	}\bibfield  {title} {\bibinfo {title} {Half-integer vs. integer effects in
			quantum synchronization of spin systems},\ }\href
	{https://doi.org/10.22331/q-2022-12-29-885} {\bibfield  {journal} {\bibinfo
			{journal} {{Quantum}}\ }\textbf {\bibinfo {volume} {6}},\ \bibinfo {pages}
		{885} (\bibinfo {year} {2022})}\BibitemShut {NoStop}%
	\bibitem [{\citenamefont {Peres}(1996)}]{PhysRevLett.77.1413}%
	\BibitemOpen
	\bibfield  {author} {\bibinfo {author} {\bibfnamefont {A.}~\bibnamefont
			{Peres}},\ }\bibfield  {title} {\bibinfo {title} {Separability criterion for
			density matrices},\ }\href {https://doi.org/10.1103/PhysRevLett.77.1413}
	{\bibfield  {journal} {\bibinfo  {journal} {Phys. Rev. Lett.}\ }\textbf
		{\bibinfo {volume} {77}},\ \bibinfo {pages} {1413} (\bibinfo {year}
		{1996})}\BibitemShut {NoStop}%
	\bibitem [{\citenamefont {Horodecki}\ \emph {et~al.}(1996)\citenamefont
		{Horodecki}, \citenamefont {Horodecki},\ and\ \citenamefont
		{Horodecki}}]{HORODECKI19961}%
	\BibitemOpen
	\bibfield  {author} {\bibinfo {author} {\bibfnamefont {M.}~\bibnamefont
			{Horodecki}}, \bibinfo {author} {\bibfnamefont {P.}~\bibnamefont
			{Horodecki}},\ and\ \bibinfo {author} {\bibfnamefont {R.}~\bibnamefont
			{Horodecki}},\ }\bibfield  {title} {\bibinfo {title} {Separability of mixed
			states: necessary and sufficient conditions},\ }\href
	{https://doi.org/https://doi.org/10.1016/S0375-9601(96)00706-2} {\bibfield
		{journal} {\bibinfo  {journal} {Physics Letters A}\ }\textbf {\bibinfo
			{volume} {223}},\ \bibinfo {pages} {1} (\bibinfo {year} {1996})}\BibitemShut
	{NoStop}%
\end{thebibliography}
\end{document}